# Sharp zero-phonon lines of single organic molecules on a hexagonal boron-nitride surface

Robert Smit[a], Arash Tebyani[a], Jil Hameury[a], Sense Jan van der Molen[a] and Michel Orrit[a,*]

[a] Huygens-Kamerlingh Onnes Laboratory, LION, Postbus 9504, 2300 RA Leiden, The Netherlands
[*] E-mail: Orrit@Physics.LeidenUniv.nl

**Abstract**

Single fluorescent molecules embedded in the bulk of host crystals have proven to be excellent probes of the dynamics in their nano environment, thanks to their narrow (about 0.1 μeV) optical linewidth of the 0-0 zero-phonon line (0-0 ZPL) at cryogenic temperatures. However, the optical linewidths of the 0-0 ZPL have been found to increase dramatically as the single molecules are located closer to a surface or interface, while no 0-0 ZPL has been detected for organic molecules on any surface. Here, we study single fluorescent terrylene molecules adsorbed on the surface of hexagonal boron-nitride (hBN) substrates. Our low-temperature results show for the first time the observation of the 0-0 ZPL of fluorescent molecules on a surface. With resonant excitation, we find 0-0 ZPL linewidths down to values that are about 10 times larger than the Fourier limit of 45 ± 3 MHz, dictated by the fluorescence lifetime. We compare our results for molecules deposited on the surfaces of annealed and non-annealed hBN flakes and we see a marked improvement in the spectral stability of the emitters after annealing. Our high-resolution spectra obtained on terrylene suggest the possibility of employing hBN in combination with a wide variety of single molecule emitters for investigation of physical phenomena at surfaces or for use in nanophotonic devices.

**Introduction**

Low-temperature spectroscopy of single-molecules, reaching near to lifetime-limited emission of the purely electronic transition (0-0 zero-phonon line or 0-0 ZPL for short)[1], has so far been exclusively performed on molecules embedded in the bulk of three-dimensional matrices. Such host matrices, for example for terrylene, can be noble gases[2], polymers[3], n-alkanes[4] (Shpol'skii matrices) or aromatics[5–7]. However, molecules that are closer to an interface or surface tend to show anomalous behavior compared to those embedded deeper in the bulk of the material. A study of single terrylene molecules in the hexadecane Shpol'skii matrix reported that molecules closer to an interface with silica were subject to a strong broadening of the 0-0 ZPL, while no 0-0 ZPL was observed for molecules located directly on the silica[8]. Likewise, a study of nanoscopic channels filled with a matrix of dibenzanthanthrene molecules in a naphthalene crystal has shown that a decrease in crystal thickness results in an increased broadening of the 0-0 ZPLs[9]. Lastly, a study on single molecules that had diffused at various depths into a polymer layer, also pointed to a broadening of the 0-0 ZPL with decreasing distance to the polymer surface[10]. Strikingly, the study claimed that no 0-0 ZPL was detected for molecules closer than 0.5 nm to the polymer surface, while molecules that were buried slightly deeper into the polymer (up to a few nm) were subject to intense broadening and rapid spectral diffusion of their 0-0 ZPL due to coupling to two-level systems (TLSs). The causes for the anomalous behavior of 0-0 ZPLs of molecules close to a surface or interface as compared to molecules in the bulk is still under debate. Possibly, high densities of TLSs at or close to the surface have lower activation energies due to the ill-condensed boundaries of a crystal. Impurities could also pile up at the crystal boundaries due to expulsion from the bulk or by adsorption from the environment and attribute to a local increase of conformational states. Despite these surface-induced effects on 0-0 ZPLs, it has been possible to reduce the dimensions of host-guest crystals and yet obtain near-lifetime-limited emission in the bulk of aromatic nanocrystals less than 100 nm thick, grown by reprecipitation from solution[11,12]. Thinner matrices for single molecules are highly desirable for applications such as single-charge sensing, which requires a close proximity of the sensing molecule to the charge[13], or potential applications in nanophotonics[14]. For the latter case, such systems-on-a-chip make use of dielectric waveguides, and a closer proximity of single molecules to waveguides is expected to lead to a better coupling to the evanescent field. First experiments with waveguide-coupled single dibenzoterrylene (DBT) molecules revealed spectral diffusion of the 0-0 ZPL attributed to charge fluctuations within the nanoguide itself[15]. With high-quality anthracene crystals of about 150 nm in thickness suspended over the waveguide channels, it was demonstrated that DBT molecules coupling to the waveguide remain stable and near-lifetime-limited for hours[16].

The advent of two-dimensional materials has initiated studies of new classes of single emitters, namely light-emitting defects and color centers[17]. Unlike the aforementioned single molecules, these defects and color centers, are covalently incorporated into the host system, and can show relatively sharp zero-phonon lines at room temperature thanks to the rigidity of their host[18]. One such promising 2D material is hexagonal Boron-Nitride (hBN), which can host emitters over a surprisingly broad spectral range, extending from the deep UV[19] up to the NIR[20], due to its large band gap of approximately 6 eV[21,22]. The atomic and electronic structure of many of these emitters still remains a puzzle and only educated guesses as to their origin have been made from quantum-chemistry models[23–26]. Contrary to these emitters inside hBN, single molecules are well-studied and have well-known chemical structures and photophysical properties. A first study at room temperature by S. Han et al.[27], showed that terrylene molecules adsorbed on hBN could be measured for a prolonged time due to a considerable decrease in bleaching rates. In the present work, we searched and found the 0-0 ZPL of terrylene molecules at low temperatures, a first-time observation for single molecules on any surface.

We present measurements of the 0-0 ZPLs of single terrylene molecules on the surface of hBN, by high-resolution excitation laser spectroscopy. We reveal remarkably narrow linewidths down to a few 100 MHz, less than a factor 10 away from the lifetime-limited linewidth of about 45 MHz. On a larger scale of GHz's up to THz's, the single terrylene molecules are subject to photo-induced spectral diffusion appearing as spectral jumps, often arising from coupling to two-level systems (TLSs). We observe clear differences between hBN substrates that were or were not annealed, prior to deposition of molecules. In particular, we see a significant reduction in the number of TLSs and in the spectral diffusion after annealing in an oxidizing atmosphere, which we attribute to the removal of (organic) contaminants from the surface of hBN. Moreover, a new spectroscopic site for terrylene is revealed after annealing with a 0-0 ZPL shift of about 20 nm to the red. Remarkably, the spectrum of molecules in this site exhibit strongly reduced vibronic couplings correlated with shifts of their intramolecular vibrations.

**Results and discussion**

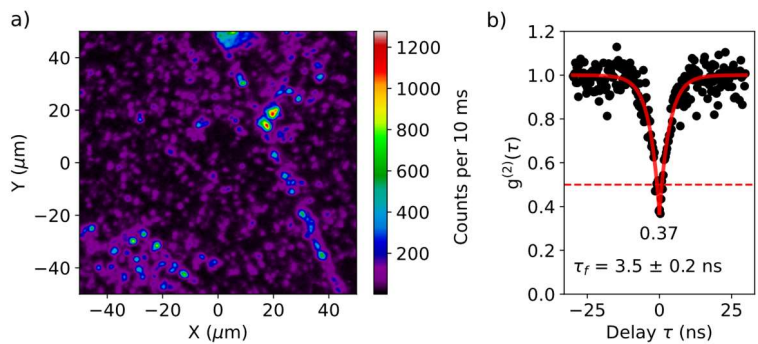

**Figure 1.** a) Confocal-fluorescence intensity map of a large (non-annealed) hBN flake showing single spots and clusters of terrylene molecules that were excited by a vibronic transition with 532 nm light. b) Antibunching recorded from a typical isolated spot. The dip does not fully extend to zero due to a relatively strong background in the fluorescence signal. This background was caused by leakage through the filter (532 nm Notch) of the high-intensity excitation light (17 kW/cm$^2$), together with a weak Raman signal from Si (520 cm$^{-1}$) and hBN (1,365 cm$^{-1}$). The red curve is a fit of the function $g^{(2)}(\tau) = 1 - ce^{-|\tau|/\tau_f}$ to the data points.

Terrylene molecules (see structure in Figure 2a) were sublimated in vacuum onto multilayer hBN flakes exfoliated onto Si/SiO$_2$ substrates (see methods section in Supplementary). The multilayer flakes were typically less than 100 nm in thickness (Figure S2.1 in Supplementary). A fluorescence map of the single emitters on hBN is shown in Figure 1a. Generally, we observe clustering of emitters around sharp lines, which are either step edges or wrinkles. Such clustering is not expected if sublimated molecules land on the hBN surface and immobilize readily. Rather, the landed molecules likely diffuse over the surface before finding a place to immobilize. Interestingly, defect emitters in exfoliated hBN are also preferentially located around sharp edges[28,29]. Isolated emitters show antibunching (Figure 1b and more in Figure S6.1 in Supplementary) with a dip below 0.5 and an average lifetime of 3.6 ± 0.2 ns. This lifetime matches well with the reported lifetime for terrylene on hBN at room temperature: 3.44 ± 0.38 ns[27]. For terrylene in general, the lifetime was found to remain constant or to increase slightly between room temperature and liquid-helium temperatures in various matrices[30].

At room temperature, the spectra of the emitters are broad (Figure S2.3, Supplementary) and generally exhibit a main emission peak around 582 nm. This clearly deviates from the observation by Han et al., who reported an emission peak around 600 nm[27]. An emission around 600 nm would correspond to a significant red-shift when compared to terrylene in organic matrices, (with the exception of *p*-dichlorobenzene, where the red-most spectroscopic site of terrylene was found at 597 nm[6]). However, by sublimating terrylene on annealed hBN flakes, we also find a small sub-population of molecules with an emission peak around 602 nm, with a spectrum that differs significantly from that of the molecules in the 582 nm site (compare Figure 2c and Figure 2d).

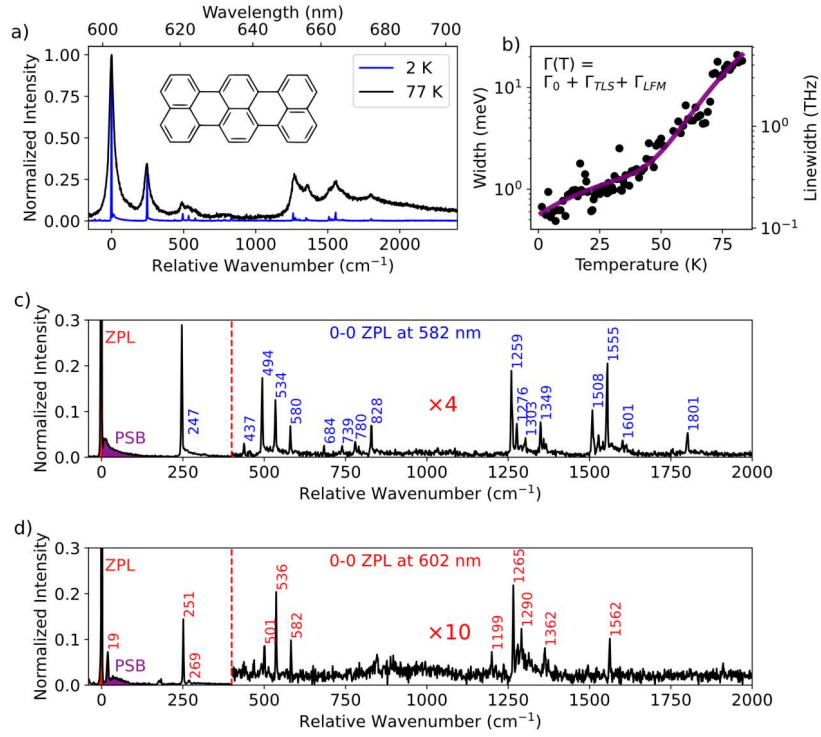

**Figure 2.** a) Emission spectra of a single terrylene molecule around 582 nm, taken at two temperatures, 2 K (blue) and 77 K (black). All spectra have been normalized to unity relative to the intensity of the 0-0 ZPL. b) Linewidth of the 0-0 ZPL of a single terrylene molecule (not the one shown in (a)). The linewidth was fitted to the equation in inset, where $\Gamma_{TLS}$ scales linearly with temperature and $\Gamma_{LFM}$ is given by the Arrhenius law, described in the main text. The linewidth at the lowest temperature was limited by the spectrometer resolution ($1.7 \pm 0.1$ cm$^{-1}$). Some spectra of this molecule can be found in section 4 of the Supplementary. c) Zoomed-in version of the spectrum at 2 K in (a) with vibrational peaks annotated by their energy in cm$^{-1}$. The part of the spectrum after 400 cm$^{-1}$ has been magnified by a factor 4. d) Details of a spectrum of a molecule with 0-0 ZPL located at 602 nm. Note that the vibronic lines are quite weak compared to those in the spectrum of (c), requiring for visibility a magnification factor of 10 for the part after 400 cm$^{-1}$. The Debye-Waller factors, which are known as the intensity of the 0-0 ZPL with respect to the combined intensity of the 0-0 ZPL (red area) and the phonon sideband (PSB, purple area), thereby excluding intramolecular vibrational peaks, are around $0.7 \pm 0.1$ for the molecule at 582 nm and $0.8 \pm 0.1$ for the molecule at 602 nm. The peak around 19 cm$^{-1}$ in (d) was included in the phonon sideband.

At liquid-helium temperatures, the spectra have narrowed down to the limit of the spectrometer's resolution ($1.7 \pm 0.1$ cm$^{-1}$). This allowed us to identify each emitter independently by its vibrational fingerprint (Figure 2a with a detailed version in Figure 2c). The spectrum in Figure 2c clearly displays the fingerprint of terrylene, which has been studied before in great detail in various matrices[31–33]. These assignments of the vibrations and comparisons of terrylene in other matrices can be found in section 3 of the Supplementary. The coarse vibrational fingerprint could also be resolved at 77 K with relative ease (Figure 2a), which makes the use of liquid helium not absolutely necessary for the identification of the emitters. In cases where the samples were prepared by spin coating of molecules in a toluene solution, instead of the sublimation method mentioned above, impurity emitters could be found regularly (section 8 in Supplementary). These emitters have 0-0 ZPLs spread over a broad range of 618 nm up to 640 nm and exhibit the fingerprint of a well-known but unidentified impurity of polymers[34] and solvents[35], such as toluene[36]. To our knowledge, no reports of this impurity on hBN have been published before, but we would like to warn readers about the possibility of wrongly assigning this impurity to intrinsic emitters of hBN.

We characterized the effect of temperature on the emission spectrum of a single terrylene molecule by the broadening of the 0-0 ZPL. We measured the linewidth or full-width-at-half-maximum for 88 intermediate temperatures in the range from 1.4 K up to 83 K. The 0-0 ZPL linewidth follows a broadening relation ($\Gamma(T) = \Gamma_0 + \Gamma_{TLS} + \Gamma_{LFM}$) that is consistent with defects in hBN[37,38] or disordered matrices, such as polymers[39]. This similarity arises from the thermal activation of tunneling two-level systems (TLSs), for which the contribution to the broadening follows a quasi-linear relationship with temperature: $\Gamma_{TLS} = bT^\alpha$, with $\alpha \approx 1$. The exponential term is accounted for by the population of quasi-localized low-frequency modes (LFMs), which follows the Arrhenius law: $\Gamma_{LFM} = w \times \exp(-E_a/k_bT)$. Here $E_a$ is defined as the activation energy of the LFMs. Although there is not much literature on the linewidth of single molecules in 3D matrices up to such high temperatures, recorded by emission spectra, one work mentions a linewidth of a single terrylene molecule in Shpol'skii matrices *n*-hexadecane and *n*-dodecane of respectively $33 \pm 3$ cm$^{-1}$ and $21 \pm 3$ cm$^{-1}$ at 50 K[40]. This compares well to a linewidth of about 25 cm$^{-1}$ at 50 K for the terrylene molecule on hBN, making this system very comparable to the bulk of a Shpol'skii matrix at higher temperatures. However, the contribution from TLSs, estimated as $21 \pm 14$ µeV/K ($5 \pm 3.5$ GHz/K) seems to be considerably larger than the typically less than 100 MHz/K for Shpol'skii matrices[39], meaning that these systems probably broaden quicker at lower temperatures. The stronger TLS broadening might be due to the closer proximity of these TLSs, as they are probably related to the (organic) contamination on the surface. Beyond the linear regime, there is a clear exponential take-off from about 50 K, which corresponds to a rather high Debye temperature of $405 \pm 41$ K. This is considerably higher than the 10-40 K mentioned for dibenzoterrylene (DBT) in anthracene[41] and found in all organic matrices. However, DBT molecules in anthracene broaden with a pure Arrhenius curve and do not suffer from additional broadening of the homogeneous linewidth due to TLSs.

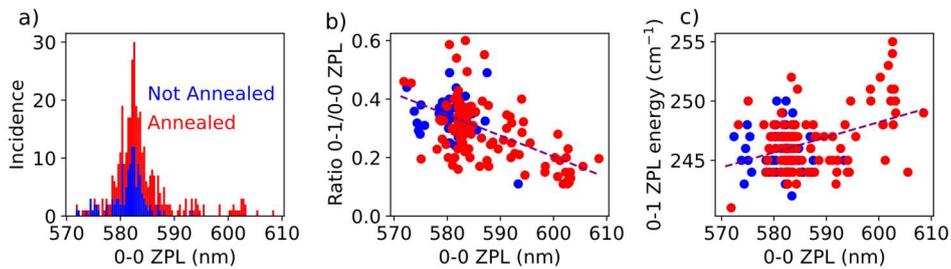

**Figure 3**. Statistics of 373 terrylene molecules divided into sets of two colors. The blue points (123 molecules) represent data on non-annealed hBN flakes, while the red points (250 molecules) correspond to flakes subject to annealing with at least 500 ˚C and for some samples up to 900 ˚C. The data in (a) shows the distribution of the 0-0 ZPL positions and indicates that after annealing a small and narrower distribution is found around 602 nm with some additional molecules between 582 nm and 602 nm. A total of 17 molecules were found with a 0-0 ZPL above 600 nm, all of them on annealed hBN. b) Shows the measured relative intensity of the main vibration (0-1 ZPL) versus the position of the 0-0 ZPL. On average, the more red-shifted molecules have a weaker vibronic coupling (Pearson's correlation coefficient r = -0.56). A less-significant relation (r = 0.4) is found for the vibration energy compared to the 0-0 ZPL position in (c).

We find the 0-0 ZPLs of terrylene molecules to be scattered over a broad range, though with clustering into specific spectroscopic sites (Figure 3a). The distribution around 582 nm – which we will call the main site – displays a relatively broad inhomogeneous broadening of about 4 nm (120 cm$^{-1}$), which is typical for disordered systems such as polyethylene[3]. On hBN flakes annealed prior to molecule deposition, we also find a new site with a relatively narrow distribution around 602 nm, which we will call the red site. Surprisingly, the molecules in this site show a significantly reduced vibronic coupling (Franck-Condon factors) between the excited state and vibronic levels of the ground state. Overall, we observe a decrease of the Franck-Condon factors when the 0-0 ZPL shifted to the red (Figure 3b). In addition, there appears to be weak a relation between the 0-0 ZPL position and the energy of the main vibration (frequency difference between the 0-1 ZPL and the 0-0 ZPL), shown in Figure 3c. We did not detect any single molecule in the red site without prior annealing of the hBN flakes. Annealing, at the temperatures we applied (500 ˚C up to 900 ˚C) is expected to remove most organic contamination[42] and to redistribute defects such as vacancies[43], as the removal of structural defects would require much higher temperatures, up to 1700 ˚C[44]. Remarkably, we find considerably fewer molecules on annealed samples, although we kept the sublimation rate of terrylene fixed or even slightly increased it (see Figure S2.2 in Supplementary). We propose as a possible explanation that terrylene anchors to (organic) contamination at the hBN surface. As annealing dramatically reduces the concentration of anchors, terrylene molecules will either leave the flake area or may aggregate to the few nucleation sites and stop fluorescing because of self-quenching. Interestingly, we find that the number of terrylene molecules on the flake increases again if prior to molecule sublimation, the annealed samples are intentionally contaminated by the spin-coating of *n*-hexadecane, which is known to form monolayers on top of hBN[45]. On the one hand, contamination could help terrylene to find more anchors and favor immobilization. On the other hand, contamination could also prevent terrylene from finding a site where it can interact strongly with the hBN, for example with some defect in hBN. We speculate that these hBN defects might be responsible for the red-shifted molecules. They would be accessible only after annealing of the hBN, right before the terrylene molecules are sublimated. The need for anchoring points could possibly be avoided by in-situ evaporation of terrylene on cold hBN surfaces, which is not possible in our setup at this moment.

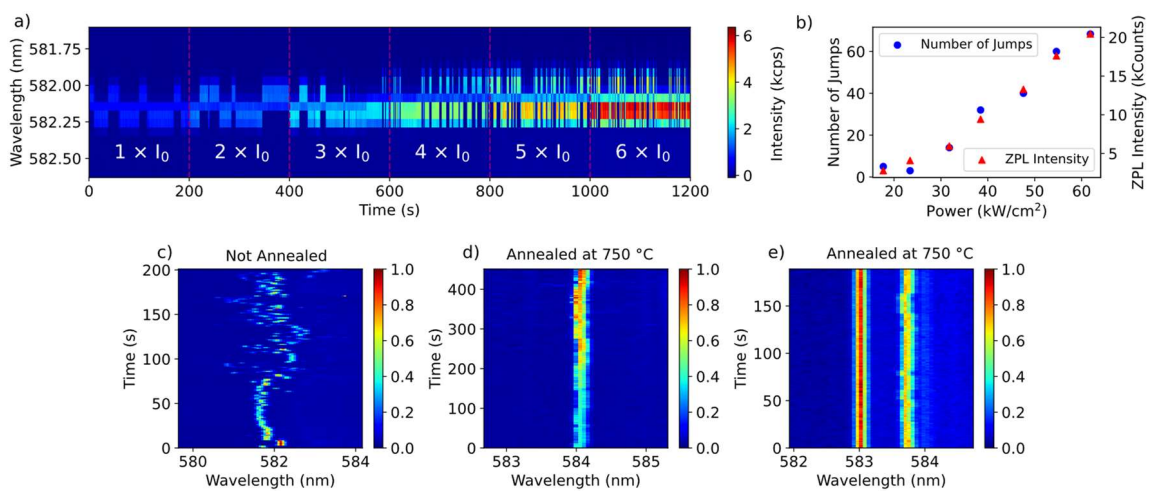

**Figure 4.** a) A 0-0 ZPL of a terrylene molecule in the main site followed for 20 minutes on an hBN flake that was annealed at 750 ˚C for 12 hours, excited by a vibronic transition with 532 nm light. The excitation intensity is raised after each frame of 200 s. The associated power densities are shown in (b), together with the number of spectral jumps that are observed in the 200 s time window and the integrated intensity of the 0-0 ZPL for a period of 1 s. Without annealing, the spectral time trace typically exhibit complex spectral diffusion as shown for example in (c). Initially, the molecule wanders around a relatively small spectral region, but extends over a much wider region from about 70 s. After annealing, the amplitude of spectral diffusion is typically strongly reduced, as in (d), even though the laser power is doubled at each 100 s interval. Some molecules do not show any spectral diffusion on the scale resolved by the spectrometer, as is the case for the leftmost molecule in (e). The spectra in (c) are measured on a different sample than the spectra in (d) and (e). More examples of spectral traces can be found in Figure S5.7 in the Supplementary.

We find that annealing of the hBN flakes improves the spectral stability of the single emitters, which we assign to jumps of (tunneling) two-level systems (TLSs). This observation is consistent with our hypothesis that these TLS's may have been located in the organic contaminants. The improvement of the spectral stability is most clear for an annealing temperature between 500 ˚C up to 750 ˚C, but much less clear for higher temperatures, up to 900 ˚C. An evident case of coupling of a molecule to a single TLS is shown in Figure 4a, which traces the 0-0 ZPL position over time at gradually-increased excitation intensities. In Figure 4b, the number of spectral jumps observed in a time window of 200 s is related to the power density of the laser spot at the position of the molecule and follows the same relation as the fluorescence intensity of the 0-0 ZPL. Therefore, these spectral jumps are laser-induced in one-photon processes. Later, we will show that spectral jumps are also observed by resonant excitation of single molecules, which requires power densities of at least three orders of magnitude weaker, due to the narrow width of the 0-0 ZPL with respect to the width of a vibronic transition.

We consider several possible sources as a cause for the spectral diffusion. One of them is a spatial jump of the molecule itself. However, rotation of the molecule can be ruled out by analysis of the fluorescence polarization (Figure S5.2, Supplementary), while super-resolution imaging did not reveal an obvious translational movement either (Figure S5.3 and S5.4, Supplementary). Clearly, the number of two-level systems is reduced significantly after annealing, while also the amplitude of spectral diffusion is observed to decrease (Figure 4a, 4d and 4e). Without annealing, the spectral jumping is in general much more complex, consisting of many, possibly coupled, levels, whose population rates could change over time (Figure 4c). The scale of the spectral jumps, extending up to a few THz in some cases, and the lack of correlation between spectral diffusion of emitters in the same focal area, points to events in the close vicinity (few nm) of the molecule. As discussed before, the molecule is likely anchored to (aggregates of) contaminants on the surface. This (nonfluorescent) contamination itself can be responsible for the spectral jumps, i.e. by (a group of) atoms tunneling between two spatial positions, perturbing the optical transition of the terrylene molecule by electrostatic or elastic dipole-dipole interaction. However, the exact nature of these TLSs remains unknown due to the random environment around the molecule.

In the data presented so far, single molecules were excited through a vibronic transition and not resonantly through the 0-0 ZPL. The 0-0 ZPL linewidth resolved in the emission spectrum is limited by the spectrometer resolution ($51 \pm 3$ GHz per pixel), while the linewidth of terrylene is expected to be up to 3 orders of magnitude narrower, around $45 \pm 3$ MHz, if it is limited by the fluorescence lifetime of $3.6 \pm 0.2$ ns. With a tunable dye laser (linewidth of a few MHz) we excited single terrylene molecules resonantly. In many cases, the molecule jumped out of resonance with the excitation laser, already in the first scan. With lower excitation intensities of a few W/cm², we could follow the molecules longer. Again, the most stable molecules were found on the annealed samples.

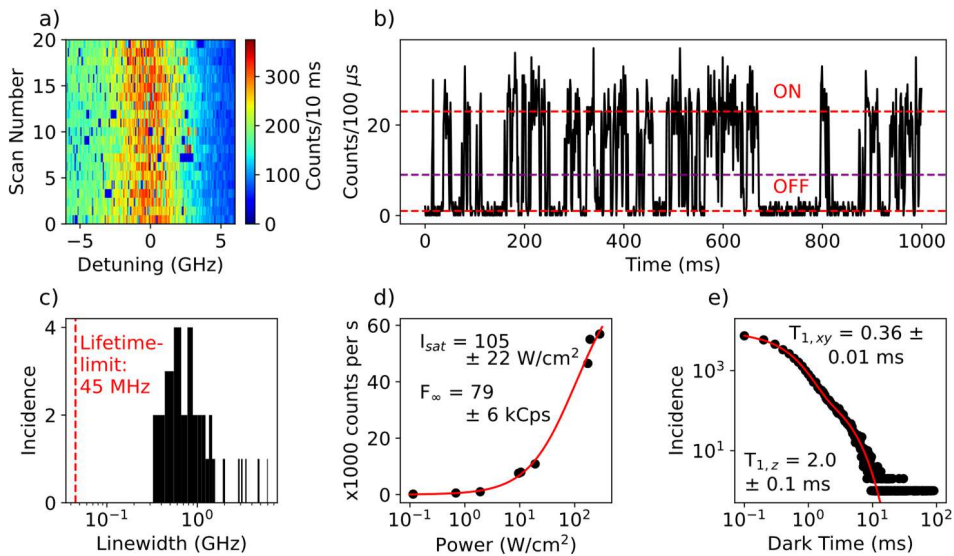

**Figure 5.** a) Excitation spectrum of the 0-0 ZPL of a single molecule that did not jump out of resonance with the laser, even at higher excitation intensities. The linewidth is about $4.7 \pm 0.3$ GHz. A resonance fluorescence time trace of the fluorescence signal of this molecule in (b) shows characteristic quantum jumps due to intersystem crossing to the triplet state. However, in some cases the molecule remained dark for much longer times than the triplet lifetime, as is also visible in (a). These long dark times may be attributed to a relatively short spectral jump back and forth between another (out-of-range) spectral position. The purple line represents the threshold set between the ON and OFF state of the fluorescence. c) Distribution of 0-0 ZPLs linewidths found for molecules on different flakes. The red dashed vertical line represents the lower limit set by the fluorescence lifetime. Due to the large distribution of linewidths, the horizontal axis has been set to a logarithmic scale. d) Saturation curve of the fluorescence of the molecule in (a). The saturation intensity is $105 \pm 22$ W/cm², 2 orders of magnitude larger than typically obtained with our setup for near-lifetime-limited emitters. This is explained by the 2 orders of magnitude broader linewidth, as the saturation intensity scales linearly with the linewidth of the transition. e) Distribution of dark periods in the resonance fluorescence signal, recorded over 60 s with 100 µs time bins (as in (b)). The characteristic timescales of the fit (red curve) are shown on the top right.

A distribution of linewidths of the 0-0 ZPL of single terrylene molecules is shown in Figure 5c. The narrowest 0-0 ZPLs are about a factor 10 above the lifetime-limited linewidth. The median of the distribution lies around 1 GHz (4.1 µeV). This is not very different from defect emitters reported for hBN, which show similar linewidths of the ZPL under resonant excitation[38]. In rare cases, we find molecules that did not move out of the scanned frequency range or only for a very short period. One example is shown in Figure 5a (others in section 7 of Supplementary). This molecule is measured on an annealed flake and has a relatively broad linewidth of $4.7 \pm 0.3$ GHz. The molecule is stable, although the asymmetric shape of the 0-0 ZPL might indicate some spectral diffusion to one side.

We show a fluorescence time trace with resonant excitation of that molecule in Figure 5b and it displays so-called quantum jumps in the fluorescence signal due to intersystem crossing to the triplet state[46]. To analyze the characteristic time scales of these quantum jumps, we determined a threshold between the ON and OFF state, where a crossing of this threshold would indicate a change from ON to OFF or vice versa. For a time trace of 60 s, the lengths of the periods where the molecule was OFF are plotted into a histogram, shown in Figure 5e. The histogram is best fitted by a bi-exponential decay, with characteristic triplet lifetimes of 360 ± 10 μs and 2.0 ± 0.1 ms. We attribute the short decay time to the indistinguishable decays of the in-plane triplet states $T_{xy}$, while the remaining long decay corresponds to the out-of-plane state $T_z$[47]. The triplet lifetimes agree well with terrylene in the extensively studied host p-terphenyl[7,46,48] and anthracene[49]. Despite the general agreement with the bi-exponential fit, the existence of datapoints at considerably longer times may point to another source of fluorescence blinking, possibly caused by relatively short-lived spectral jumps. We continued by recording an antibunching histogram of the resonance fluorescence (Figure S7.5 in Supplementary). No Rabi oscillations are observed in the histogram, which is expected when the linewidth is significantly broadened by dephasing. The measured linewidth of 4.7 ± 0.3 GHz would correspond to a lower bound of the decoherence time $T_2^*$ of about 68 ± 4 ps.

**Conclusion**

We have shown that terrylene molecules adsorbed on the surface of hBN become narrow emitters at low temperature, with linewidths as narrow as a few 100 MHz up to a few GHz, which is similar to intrinsic hBN defects[37,38,50,51]. Their relative spectral stability, in contrast to single molecules adsorbed on any other surface so far, made it possible to observe 0-0 ZPLs on a surface for the first time. Moreover, we have found a way to considerably improve the spectral stability by annealing the hBN substrates before the molecules were deposited, which points to (organic) contamination as a potential source of the spectral jumps. Spectral instabilities for emitters hosted by hBN are not new and are also observed for defect emitters inside hBN, which become particularly prominent at low temperature[24,37,38,38,51–53]. In future experiments, a well-known system such as terrylene could help shed light on these various issues of spectral instabilities and dephasing mechanisms, in order to improve the quality of emitters hosted by hBN. Moreover, the sensitivity of the molecules to their surroundings shows that our results are relevant for the study of molecule-surface interactions as well as the study of physical phenomena taking place at the surface of hBN.

To our surprise, we also found a new spectral site after annealing, with a significantly altered spectrum compared to molecules in the main site. The shifts of the vibronic lines and their weak vibronic coupling could be the result of strong interaction with hBN, possibly with some defects. We hypothesize that the reduced amount of organic contamination frees these adsorption sites for single terrylene molecules.

Despite the rigidity of the hBN host, the molecules do not benefit yet from a narrower 0-0 ZPL at higher temperatures. Contrary to emitting defects, which can be buried deep inside the multilayer hBN, the single molecules are likely much more sensitive to dynamics at the surface, leading to a stronger broadening with temperature by the activation of two-level systems. The compatibility of single molecules and hBN paves the way for further studies that could implement the encapsulation of single molecules between hBN layers. We anticipate that the passivation of the molecule's surrounding by one or more hBN layer(s) could further improve their spectral stability and could help bring their 0-0 ZPL linewidth closer to the lifetime limit, even at temperatures higher than those of liquid-helium.

**Acknowledgements**

We thank research engineer Harmen van der Meer for his help with the repair of our cryostat. Furthermore, we acknowledge the AFM Facility of the Leiden Institute of Physics (LION) for enabling the AFM measurements. We also thank in particular Dr. Zoran Ristanovic and Dr. Amin Moradi and all other past group members that have contributed to the build-up of our setups, which have made this work possible. Lastly, we are grateful for the funding of this research, which was provided by the NWO (Spinoza prize 2017).

**References**

1. Wang, Y., Bushmakin, V., Stein, G. A., Schell, A. W. & Gerhardt, I. Optical Ramsey spectroscopy on a single molecule. *Optica* **9**, 374–378 (2022).
2. Deperasińska, I., Kozankiewicz, B., Biktchantaev, I. & Sepioł, J. Anomalous Fluorescence of Terrylene in Neon Matrix. *J. Phys. Chem. A* **105**, 810–814 (2001).
3. Tchénio, P., Myers, A. B. & Moerner, W. E. Optical studies of single terrylene molecules in polyethylene. *J. Lumin.* **56**, 1–14 (1993).
4. Plakhotnik, T. & Donley, E. A. Statistics of a single terrylene molecule in hexadecane. *J. Lumin.* **86**, 175–180 (2000).
5. Gorshelev, A. A. *et al.* Ortho-Dichlorobenzene Doped with Terrylene—a Highly Photo-Stable Single-Molecule System Promising for Photonics Applications. *ChemPhysChem* **11**, 182–187 (2010).
6. Navarro, P., Tian, Y., van Stee, M. & Orrit, M. Stable single-molecule lines of terrylene in polycrystalline para-dichlorobenzene at 1.5 K. *Chemphyschem Eur. J. Chem. Phys. Phys. Chem.* **15**, 3032–3039 (2014).
7. Kummer, S., Basché, Th. & Bräuchle, C. Terrylene in p-terphenyl: a novel single crystalline system for single molecule spectroscopy at low temperatures. *Chem. Phys. Lett.* **229**, 309–316 (1994).
8. Fleury, L., Gruber, A., Dräbenstedt, A., Wrachtrup, J. & von Borczyskowski, C. Low-Temperature Confocal Microscopy on Individual Molecules near a Surface. *J. Phys. Chem. B* **101**, 7933–7938 (1997).
9. Gmeiner, B., Maser, A., Utikal, T., Götzinger, S. & Sandoghdar, V. Spectroscopy and microscopy of single molecules in nanoscopic channels: spectral behavior vs. confinement depth. *Phys. Chem. Chem. Phys.* **18**, 19588–19594 (2016).
10. Vainer, Y. G., Sobolev, Y. I., Naumov, A. V., Osad'ko, I. S. & Kador, L. Fluorescence microscopy and spectroscopy of subsurface layer dynamics of polymers with nanometer resolution in the axial direction. *Faraday Discuss.* **184**, 237–249 (2015).
11. Pazzagli, S. *et al.* Self-Assembled Nanocrystals of Polycyclic Aromatic Hydrocarbons Show Photostable Single-Photon Emission. *ACS Nano* **12**, 4295–4303 (2018).
12. Schofield, R. C. *et al.* Narrow and Stable Single Photon Emission from Dibenzoterrylene in para-Terphenyl Nanocrystals. *ChemPhysChem* **23**, e202100809 (2022).
13. Faez, S., van der Molen, S. J. & Orrit, M. Optical tracing of multiple charges in single-electron devices. *Phys. Rev. B* **90**, 205405 (2014).
14. Toninelli, C. *et al.* Single organic molecules for photonic quantum technologies. *Nat. Mater.* (2021) doi:10.1038/s41563-021-00987-4.
15. Shkarin, A. *et al.* Nanoscopic Charge Fluctuations in a Gallium Phosphide Waveguide Measured by Single Molecules. *Phys. Rev. Lett.* **126**, 133602 (2021).
16. Ren, P. *et al.* Photonic-circuited resonance fluorescence of single molecules with an ultrastable lifetime-limited transition. *Nat. Commun.* **13**, 3982 (2022).

17. Tran, T. T., Bray, K., Ford, M. J., Toth, M. & Aharonovich, I. Quantum emission from hexagonal boron nitride monolayers. *Nat. Nanotechnol.* **11**, 37–41 (2016).
18. Grosso, G. *et al.* Tunable and high-purity room temperature single-photon emission from atomic defects in hexagonal boron nitride. *Nat. Commun.* **8**, 705 (2017).
19. Bourrellier, R. *et al.* Bright UV Single Photon Emission at Point Defects in h-BN. *Nano Lett.* **16**, 4317–4321 (2016).
20. Camphausen, R. *et al.* Observation of near-infrared sub-Poissonian photon emission in hexagonal boron nitride at room temperature. *APL Photonics* **5**, 076103 (2020).
21. Tarrio, C. & Schnatterly, S. E. Interband transitions, plasmons, and dispersion in hexagonal boron nitride. *Phys. Rev. B* **40**, 7852–7859 (1989).
22. Cassabois, G., Valvin, P. & Gil, B. Hexagonal boron nitride is an indirect bandgap semiconductor. *Nat. Photonics* **10**, 262–266 (2016).
23. Gu, R. *et al.* Engineering and Microscopic Mechanism of Quantum Emitters Induced by Heavy Ions in hBN. *ACS Photonics* **8**, 2912–2922 (2021).
24. Li, X. *et al.* Nonmagnetic Quantum Emitters in Boron Nitride with Ultranarrow and Sideband-Free Emission Spectra. *ACS Nano* **11**, 6652–6660 (2017).
25. Aharonovich, I., Tetienne, J.-P. & Toth, M. Quantum Emitters in Hexagonal Boron Nitride. *Nano Lett.* **22**, 9227–9235 (2022).
26. Dev, P. Fingerprinting quantum emitters in hexagonal boron nitride using strain. *Phys. Rev. Res.* **2**, 022050 (2020).
27. Han, S. *et al.* Photostable fluorescent molecules on layered hexagonal boron nitride: Ideal single-photon sources at room temperature. *J. Chem. Phys.* **155**, 244301 (2021).
28. Choi, S. *et al.* Engineering and Localization of Quantum Emitters in Large Hexagonal Boron Nitride Layers. *ACS Appl. Mater. Interfaces* **8**, 29642–29648 (2016).
29. Yim, D., Yu, M., Noh, G., Lee, J. & Seo, H. Polarization and Localization of Single-Photon Emitters in Hexagonal Boron Nitride Wrinkles. *ACS Appl. Mater. Interfaces* **12**, 36362–36369 (2020).
30. Harms, G. S., Irngartinger, T., Reiss, D., Renn, A. & Wild, U. P. Fluorescence lifetimes of terrylene in solid matrices. *Chem. Phys. Lett.* **313**, 533–538 (1999).
31. Navarro, P. *et al.* Electron Energy Loss of Terrylene Deposited on Au(111): Vibrational and Electronic Spectroscopy. *J. Phys. Chem. C* **119**, 277–283 (2015).
32. Tchénio, P., Myers, A. B. & Moerner, W. E. Vibrational analysis of the dispersed fluorescence from single molecules of terrylene in polyethylene. *Chem. Phys. Lett.* **213**, 325–332 (1993).
33. Kummer, S. *et al.* Absorption, excitation, and emission spectroscopy of terrylene in p-terphenyl: Bulk measurements and single molecule studies. *J. Chem. Phys.* **107**, 7673–7684 (1997).
34. Fleury, L. *et al.* Single Molecule Spectra of an Impurity Found in N-Hexadecane and Polyethylene. *Mol. Cryst. Liq. Cryst. Sci. Technol. Sect. Mol. Cryst. Liq. Cryst.* **283**, 81–87 (1996).
35. Neumann, A., Lindlau, J., Thoms, S., Basché, T. & Högele, A. Accidental Contamination of Substrates and Polymer Films by Organic Quantum Emitters. *Nano Lett.* **19**, 3207–3213 (2019).
36. Trinh, C. T., Lee, J. & Lee, K.-G. Fluorescent impurity emitter in toluene and its photon emission properties. *Sci. Rep.* **8**, 8273 (2018).
37. Sontheimer, B. *et al.* Photodynamics of quantum emitters in hexagonal boron nitride revealed by low-temperature spectroscopy. *Phys. Rev. B* **96**, 121202 (2017).
38. White, S. *et al.* Phonon dephasing and spectral diffusion of quantum emitters in hexagonal boron nitride. *Optica* **8**, 1153 (2021).
39. Vainer, Yu. G., Naumov, A. V., Kol'chenko, M. A. & Personov, R. I. Quasi-localized low-frequency vibrational modes of disordered solids I. Study by photon echo. *Phys. Status Solidi B* **241**, 3480–3486 (2004).
40. Banasiewicz, M., Wiącek, D. & Kozankiewicz, B. Photo-bleaching of single terrylene molecules in Shpol'skii matrices at elevated temperatures. *Chem. Phys. Lett.* **425**, 289–293 (2006).
41. Nicolet, A. A. L. *et al.* Single Dibenzoterrylene Molecules in an Anthracene Crystal: Main Insertion Sites. *ChemPhysChem* **8**, 1929–1936 (2007).
42. Li, C. *et al.* Purification of single-photon emission from hBN using post-processing treatments. *Nanophotonics* **8**, 2049–2055 (2019).
43. Zobelli, A., Ewels, C. P., Gloter, A. & Seifert, G. Vacancy migration in hexagonal boron nitride. *Phys. Rev. B* **75**, 094104 (2007).

44. Chen, X. *et al.* The effects of post-annealing technology on crystalline quality and properties of hexagonal boron nitride films deposited on sapphire substrates. *Vacuum* **199**, 110935 (2022).
45. Arnold, T., Forster, M., Fragkoulis, A. A. & Parker, J. E. Structure of Normal-Alkanes Adsorbed on Hexagonal-Boron Nitride. *ACS Publications* (2014) doi:10.1021/jp4063059.
46. Basché, T., Kummer, S. & Bräuchle, C. Direct spectroscopic observation of quantum jumps of a single molecule. *Nature* **373**, 132–134 (1995).
47. Lawetz, V., Orlandi, G. & Siebrand, W. Theory of Intersystem Crossing in Aromatic Hydrocarbons. *J. Chem. Phys.* **56**, 4058–4072 (1972).
48. Banasiewicz, M., Morawski, O., Wiącek, D. & Kozankiewicz, B. Triplet population and depopulation rates of single terrylene molecules in p-terphenyl crystal. *Chem. Phys. Lett.* **414**, 374–377 (2005).
49. Nicolet, A., Kol'chenko, M. A., Kozankiewicz, B. & Orrit, M. Intermolecular intersystem crossing in single-molecule spectroscopy: Terrylene in anthracene crystal. *J. Chem. Phys.* **124**, 164711 (2006).
50. Tran, T. T. *et al.* Resonant Excitation of Quantum Emitters in Hexagonal Boron Nitride. *ACS Photonics* **5**, 295–300 (2018).
51. Konthasinghe, K. *et al.* Rabi oscillations and resonance fluorescence from a single hexagonal boron nitride quantum emitter. *Optica* **6**, 542 (2019).
52. Ngoc My Duong, H. *et al.* Effects of High-Energy Electron Irradiation on Quantum Emitters in Hexagonal Boron Nitride. *ACS Appl. Mater. Interfaces* **10**, 24886–24891 (2018).
53. Fournier, C. *et al.* Position-controlled quantum emitters with reproducible emission wavelength in hexagonal boron nitride. *Nat. Commun.* **12**, 3779 (2021).

# Sharp zero-phonon lines of single organic molecules on a hexagonal boron-nitride surface

Robert Smit[a], Arash Tebyani[a], Jil Hameury[a], Sense Jan van der Molen[a] and Michel Orrit[a,*]

[a] Huygens-Kamerlingh Onnes Laboratory, LION, Postbus 9504, 2300 RA Leiden, The Netherlands
[*] E-mail: Orrit@Physics.LeidenUniv.nl

The supporting information contains additional data that was not included in the main text. In the main text, we frequently refer to the various sections and figures that are part of this document.

**S1. Methods**

Flakes of hBN from a single crystal (HQ Graphene) were transferred to the substrate by the exfoliation method, where the layers are cleaved using scotch tape. The substrate for hBN was a Si wafer (University Wafer) coated with a 300 nm oxide layer. In early experiments, the exfoliated samples were cleaned in acetone to remove any residue left from the tape. In later experiments, the cleaning step after exfoliation was omitted, as it might rather contaminate than clean the freshly cleaved hBN flakes. Optical inspection showed that the multilayer flakes were present on the substrate and their lateral sizes varied from a few μm up to a few 100 μm. Annealing of the hBN was performed in a tube oven (Thermcraft) in a moderate vacuum of about $10^{-2}$ mbar of residual air.

Terrylene (synthesized by Mercachem) crystals were placed on the bottom of a sublimation apparatus (Figure S1.1). The substrate with the hBN flakes was suspended by carbon tape and was in thermal contact with the cold finger that was cooled by water ice. The atmosphere was pumped to vacuum and the flask was heated on a hot plate for a sublimation time of 5 minutes. A heating temperature of the terrylene crystals between 120 ˚C and 140 ˚C was found to yield a suitable concentration of terrylene molecules on hBN (Figure 1a in main text). Alternatively to the sublimation method, we also employed spin coating as a method for the deposition of molecules. In that case, terrylene was dissolved in toluene (Acros Organics, 99.85%) and diluted to a concentration below 1 nmol/mol. A few droplets of the terrylene solution (around 10-50 μL, depending on substrate size) were pipetted onto the substrate. Spin-coating followed at 2000 rpm for 20 s and was terminated by a drying step at 4000 rpm for an additional 20 s.

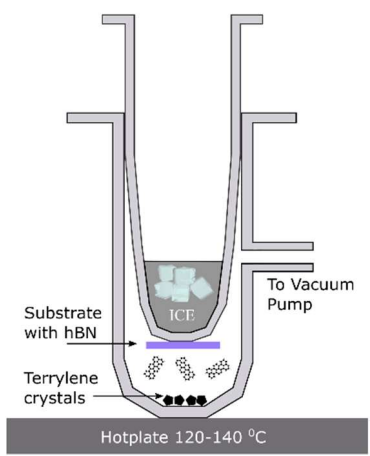

**Figure S1.1.** a) Schematic representation of the sublimation apparatus, which consists of two round-bottom flasks, where the top one is inserted into the bottom one, sealed by vacuum grease (Apiezon) and pumped by a vacuum pump from the side. The bottom of the flask is heated on a hot plate to approximately 120-140 °C. The crystals are placed inside, in contact with the heated bottom. The sample is fixed with carbon tape on the cold finger and cooled with water ice.

The samples were fixed to a sample holder and inserted into a flow cryostat (Janis SVT-200-5) that can cool down to 1.2 K. Before cooling down, the cryostat was purged three times by pumping out all the gases inside and exchanging them for dry nitrogen gas. The purging did not seem to have any noticeable effect on the exposed terrylene molecules. The cryostat contains an objective (0.85 NA, Edmund Optics) that is immersed in liquid helium and forms part of a home-built confocal setup. For the spectroscopy experiments, two excitation sources were used. For a relatively broadband excitation, a 532 nm laser (Sprout G-15W, Lighthouse Photonics) was used, which is phase-locked with six longitudinal modes spanning over 4.25 GHz. As a narrow and tunable excitation source (approximately 1 MHz linewidth) a Coherent 699 dye ring laser was used, which was operated with Rhodamine 6G dye and pumped by a Coherent Verdi V2 laser (5 W at 532 nm). The wavelength of the dye laser was monitored with MHz precision using a High Finesse WS6-200 wavemeter. The emission spectra were recorded using a Horiba iHR320 spectrometer, which was coupled to a liquid-nitrogen-cooled Symphony II CCD detector. The spectrometer can be operated with three diffraction gratings with 150, 600 and 1200 lines/mm, yielding a spectral resolution down to $1.7 \pm 0.1$ cm$^{-1}$ around a wavelength of $580 \pm 10$ nm. Confocal fluorescence imaging was performed using a scanning mirror (Newport, FSM-300-01) and fluorescence was detected with one or two APDs (Excelitas, SPCM-AQRH-16). The signal from the two APDs was time-correlated using a PicoHarp 300 from PicoQuant in combination with a PHR-800 router. A programmable delay was set on the stop channel by a delay box (Ortec DB463).

## S2. Room temperature measurements

*AFM profiles*

AFM characterization of the exfoliated hBN flakes shows that typical thicknesses are spread out over a broad range from 40 nm up to more than 300 nm, though in general they are below 100 nm in thickness. Planar sizes are typically at least 10 µm x 10 µm, but never more than a few hundred µm in diameter. In some cases, wrinkles can be observed in the sheet (Figure S2.1d). Figures S2.1e and S2.1f show the same flake before and after the annealing process (500 ˚C for 12 h). Most noticeable is that the contamination/tape residue outside of the flake area are efficiently removed by the annealing process.

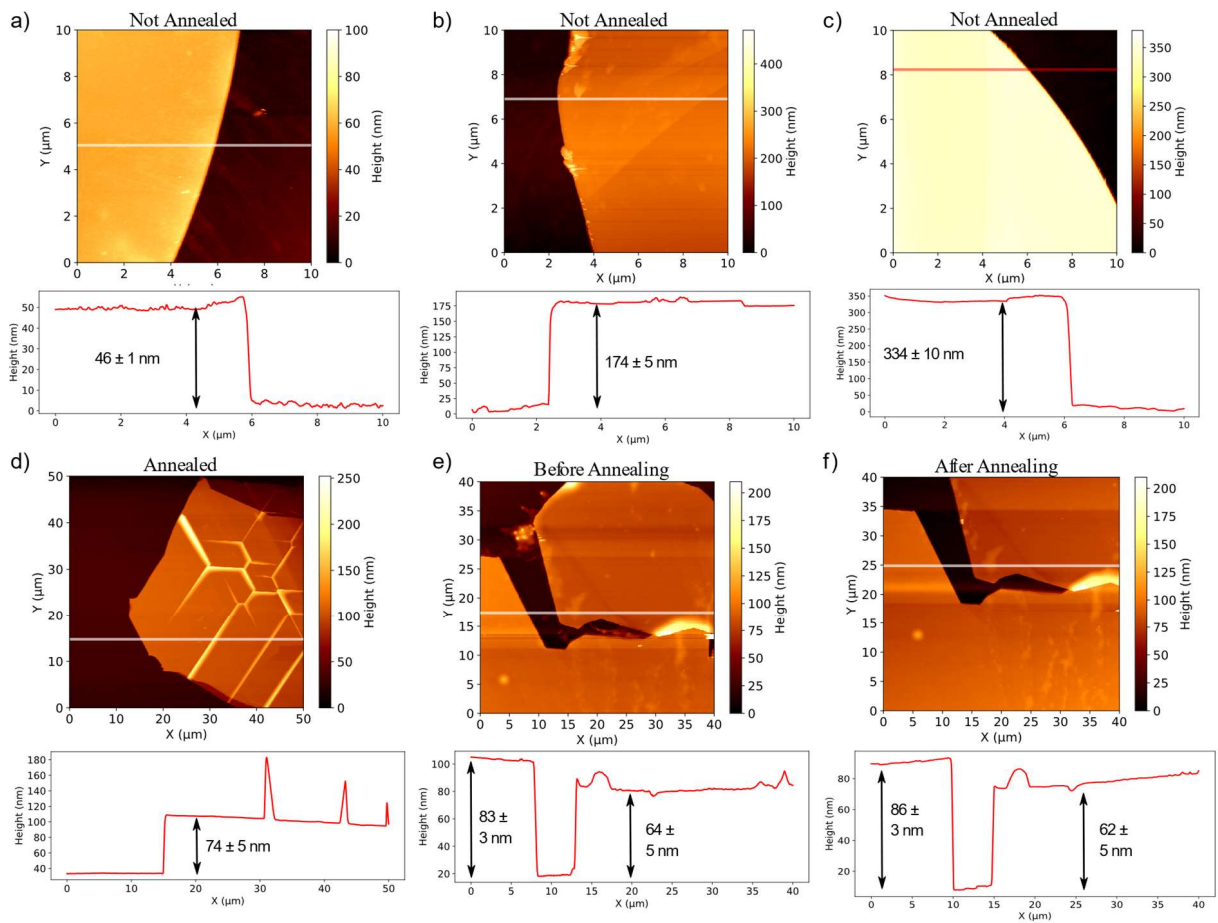

**Figure S2.1.** a,b,c) AFM scans of a small part (10x10 µm$^2$) of non-annealed exfoliated hBN flakes on a Si/SiO$_2$ substrate with under each 2D map the corresponding height profile along the white line (red in c). The thickness of hBN flakes was found to vary significantly from flake to flake, ranging from 46 up to 334 nm, but typically below 100 nm. The flake in (d) was annealed at 500 ˚C for 12 hours. The AFM images in (e) and (f) show the same flake, respectively before and after annealing at 500 ˚C for 12 hours. Tape residues or other contaminants on the substrate, outside of the flakes, are efficiently removed by annealing. On the flakes themselves there is no significant difference. Possibly some contamination on the substrate in the bottom right corner, protected by the hBN flake, does not get removed by annealing. The AFM image in (d) shows clear wrinkles in the hBN sheet, appearing as tall peaks in the AFM profile.

The fluorescence image in Figure S2.2a shows the hBN flake measured by AFM before and after annealing (Figures S2.1e, f). We deposited terrylene on this flake after annealing by sublimation at an increased rate, by heating to 135-140 ˚C for 5 minutes. Surprisingly, the number of visible emitters is much lower than for the non-annealed flake in Figure 1a of the main text. We systematically find less molecules on the annealed flakes, than on non-annealed flakes. Although there are fewer molecules, there are some locations with higher densities of emitters, which seem to have more surface roughness (Figure S2.2b). Fewer emitters are typically found on the more pristine parts of the flake, which can be near-to atomically flat (Figure S2.3).

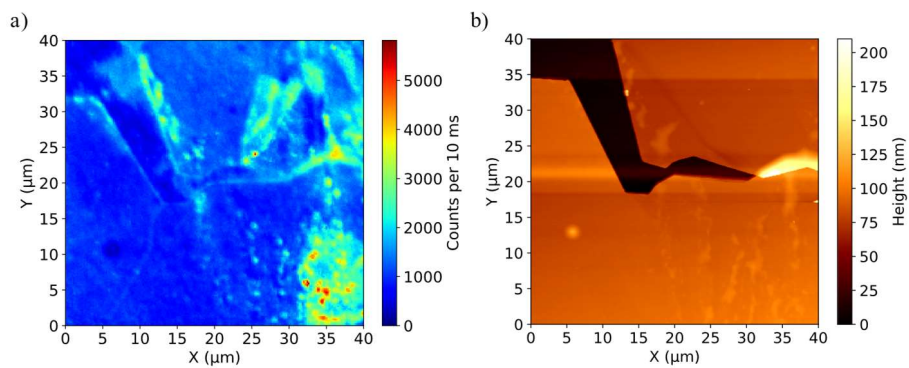

**Figure S2.2**. a) Fluorescence image of the same area that we measured with AFM in Figure S2.1e, f, before and after annealing. For comparison, the latter AFM image, after annealing, is recalled in b. The flake was annealed at 500 ˚C for 12 h. Even though we increased the sublimation temperature to 135-140 ˚C, the number of visible emitters is clearly much less than for example in the non-annealed flake in Figure 1a of the main text. Terrylene molecules are mostly found in the bottom right corner of the image. In this region, the AFM profile shows more surface roughness, which might be caused by contamination under the hBN layer. The flatter regions of the hBN flake show less fluorescence. The AFM measurement was done before deposition of the molecules and subsequent low-temperature experiments. The region in (a) for 25 < X < 30 μm and 23 < Y < 35 μm appears to have changed with respect to the same region in b), perhaps due to a folding of a part of the flake by the AFM tip.

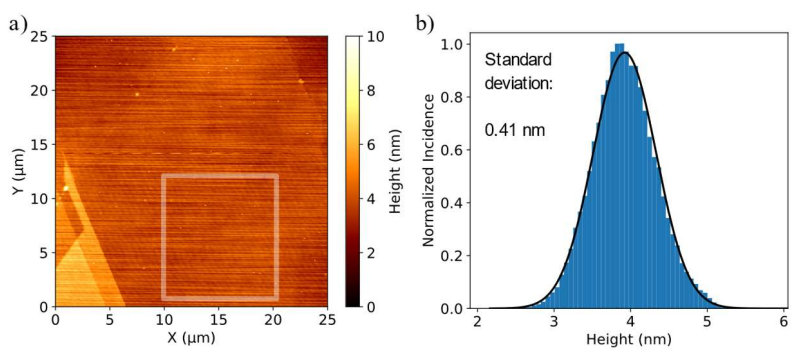

**Figure S2.3**. a) AFM image of a flat region on an annealed hBN flake (500 ˚C for 12 h). The white rectangle encloses the region that builds up the histogram of surface heights in b). The Gaussian distribution of surface heights shows a standard deviation of about 0.41 nm. Tilts and drifts in the scan were subtracted by a line-by-line linear fit.

*Room temperature spectra of terrylene*

The broadness of the spectra of single emitters on hBN at room temperature makes it difficult to identify their origin, due to a lack of spectral features. Moreover, typical defect emitters in hBN contain an emission band shifted about 1,365 $cm^{-1}$ from the main emission peak. This is due to the emitter's coupling to the B-N stretch vibrational mode, which is Raman active as shown in Figure S2.4. The B-N stretch is very close to the C-C stretch modes of polyaromatic hydrocarbons, which typically lie around 1,200-1,300 $cm^{-1}$. Hence, the distinction between various emitters present in hBN itself, impurities of the toluene and finally the terrylene itself was inherently difficult.

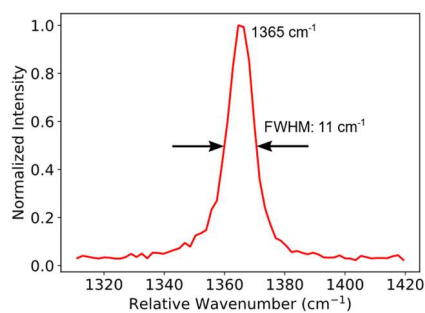

**Figure S2.4**. Spectrum of the inelastically scattered light from a multilayer hBN flake with a 532 nm excitation wavelength. The single Raman peak observed at 1365 $cm^{-1}$ with a FWHM of 11 $cm^{-1}$, is significantly broader than the spectrograph resolution, around 1-2 $cm^{-1}$. The spectrum is obtained at room temperature with a 1200 lines/mm grating and a 0.1 mm slit size. The position and width of the Raman peak conforms with the data provided by the supplier hq-graphene.

For terrylene in general, we find molecules whose spectra are broadened in such a way that no single vibrations can be resolved, as different modes merge into broad bands. A common spectrum of a terrylene molecule on hBN looks like the spectrum in Figure S2.5a. The strongest peak around 580 nm is a superposition of the phonon-broadened purely electronic transition and of the main low-frequency vibration around 245-250 cm$^{-1}$. The second band around 630 nm is dominated by a bunch of C-C stretch modes with first overtones around 680 nm. On some of the annealed samples, we find molecules that are less broadened, such as in Figure S2.5b. In this spectrum, a more detailed structure of the vibrations is revealed with a higher-resolution grating (600 lines/mm). Consistent with the low temperature results, we find most of the molecules to peak around 580 nm.

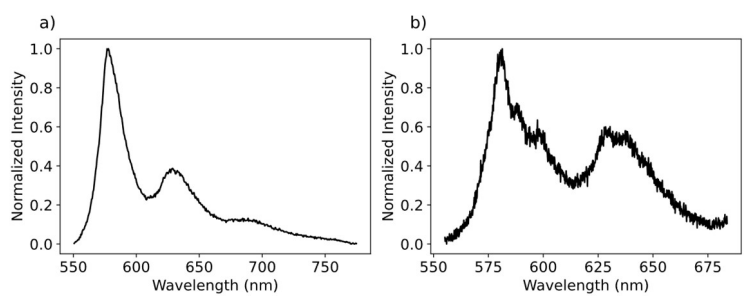

**Figure S2.5.** Emission spectra of terrylene on hBN at room temperature, recorded with a 150 lines/mm grating for the spectrum in a) and a 600 lines/mm grating in b). The spectra could stem from ensembles of molecules ($g^{(2)}(\tau)$ was not measured), which may lead to additional inhomogeneous broadening.

*Adsorption of terrylene on hBN*

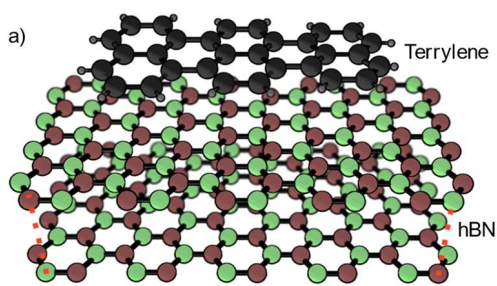

**Figure S2.6.** Schematic perspective view of terrylene on hBN, with brown colors displayed as boron and green as nitrogen. As the real adsorption sites for terrylene on hBN are unknown, we assume that, similar to pentacene at monolayer coverage[1,2], terrylene lies flat on the hBN substrate, bonded by weak Van der Waals forces. The hBN monolayers stack with boron facing nitrogen atoms at adjacent planes (dashed red lines).

Terrylene in isolated form has a practically flat structure (see Figure S2.6). In some matrices terrylene was found to exhibit a slight twist of a few degrees of the outer naphthalene units around the central naphthalene unit according to quantum chemistry calculations[3,4]. The interatomic distance of nitrogen and boron in the hexagonal lattice of hBN is 1.45 Å[5] and matches well within about 7% of the C-C bond lengths of the terrylene molecule[6]. In addition, the interlayer distance between monolayer sheets, about 3 Å, is practically identical in graphite and hBN crystals[7]. Therefore we expect the terrylene molecule, itself in approximation a small patch of graphene terminated by hydrogen atoms, to lie flat on the surface of hBN, bound by weak Van der Waals interactions, which are also responsible for the stacking of hBN or graphene monolayers and even hBN/graphene heterostructures[8]. Pentacene, a molecule comparable to terrylene, was indeed found to lie flat on the surface of graphene[9] and of hBN as a monolayer[1,2].

Unlike in graphene, the alternating arrangement of nitrogen and boron in hBN opens up a band gap, which was measured to be 6.08 eV for the bulk crystal using two-photon excitation spectroscopy[10]. The large band gap prevents the exchange of electrons or energy with terrylene, which has a HOMO-LUMO gap between the singlet ground and excited state in the range of 2.1 ± 0.1 eV[11]. In fact, the band gap of hBN is sufficiently large to probably prevent any exchange of electrons or energy with basically all known fluorescent dyes used in single-molecule spectroscopy.

## S3. Comparison of terrylene spectra on hBN and other matrices

Below we list the vibrational frequencies measured in Figure 2c and 2d in the main text with an assignment of the specific vibrations as reported in the literature and a comparison to terrylene in a few three-dimensional matrices. The positions of the vibrational frequencies and the number of lines may however vary slightly from molecule to molecule.

**Table S3.1.** Frequencies (in cm$^{-1}$) and assignments of the vibrations of the terrylene molecules in the main site at 582 nm and the red site at 602 nm, as shown in Figure 2c and 2d in the main text. The vibrational frequencies are compared with terrylene measured in three types of matrices, respectively a semi-crystalline polymer, a Shpol'skii matrix and an aromatic matrix.

| Assignment[6,12] | hBN (main site): 582 nm | hBN (red site): 602 nm | Polyethylene[12] | n-Hexadecane[3] | Anthracene[13] |
|---|---|---|---|---|---|
| 0-0 Zero-phonon line | 0 | 0 | 0 | 0 | 0 |
| Long-axis stretching | 246 | 251 | 242 | 241 | 247 |
| | 437 | 439? (very weak) | 442 | 439 | |
| 2 × 246 | 494 | 501 | 487 | | 496 |
| In-plane ring deformation | 534 | 536 | 534 | 532 | 536 |
| | 580 | 582 | 584 | 581 | 584 |
| | 684 | | | | |
| 3 × 246 | 739 | | 736 | | |
| 534 + 246 | 780 | | 780 | | |
| | 828 | 847? | 830 | | 843 |
| Aromatic C=C stretch | 1259 | 1265 | 1269 | 1269 | |
| | 1276 | | 1280 | 1279 | 1278 |
| | 1303 | 1290 | 1297 | 1309 | 1292 |
| | 1349 | | 1355 | 1357 | |
| | 1359 | | | | |
| | 1366 | 1362 | | | 1366 |
| 1259 + 246 | 1508 | | 1529 | 1506 | |
| 1276 + 246 | 1527 | | | 1522 | |
| | 1555 | 1562 | 1556 | 1553 | 1566 |
| | 1601 | | 1580 | | |
| 1555 + 246 | 1801 | | 1802 | | 1815 |

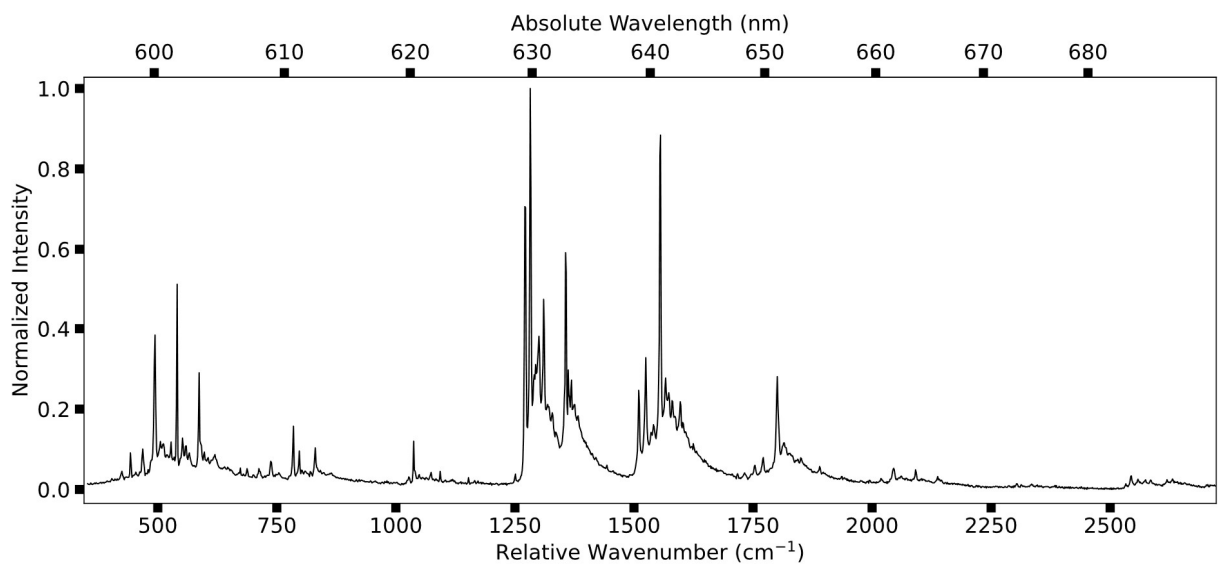

**Figure S3.1.** Emission spectrum from the resonantly-excited molecule in Figure 5 in the main text. The excitation wavelength was 582.38 nm and we used a 590 Long pass filter to block the laser light. Consequently, the 0-0 ZPL and main vibrational peak is not visible. The spectrum was recorded for 150 s using a 1,200 lines/mm grating and a 50 μm slit size. T = 2 K.

## S4. Thermal broadening of the 0-0 ZPL linewidth

We measured the 0-0 ZPL linewidths in Figure 2b in the main text by recording the spectra of a single molecule located on an annealed hBN flake (750 °C for 12 h). A subset of 9 out of the 88 different temperatures at which we recorded spectra, is shown in Figure S4.1. A weak contribution from another emitter around 579-580 nm is observed in the spectrum at 2 K. The relative intensities of the two emitters varied due to drifts in the laser spot as the temperature changed. To account for these drifts, we realigned the microscope to the main molecule at intervals of about 5 K.

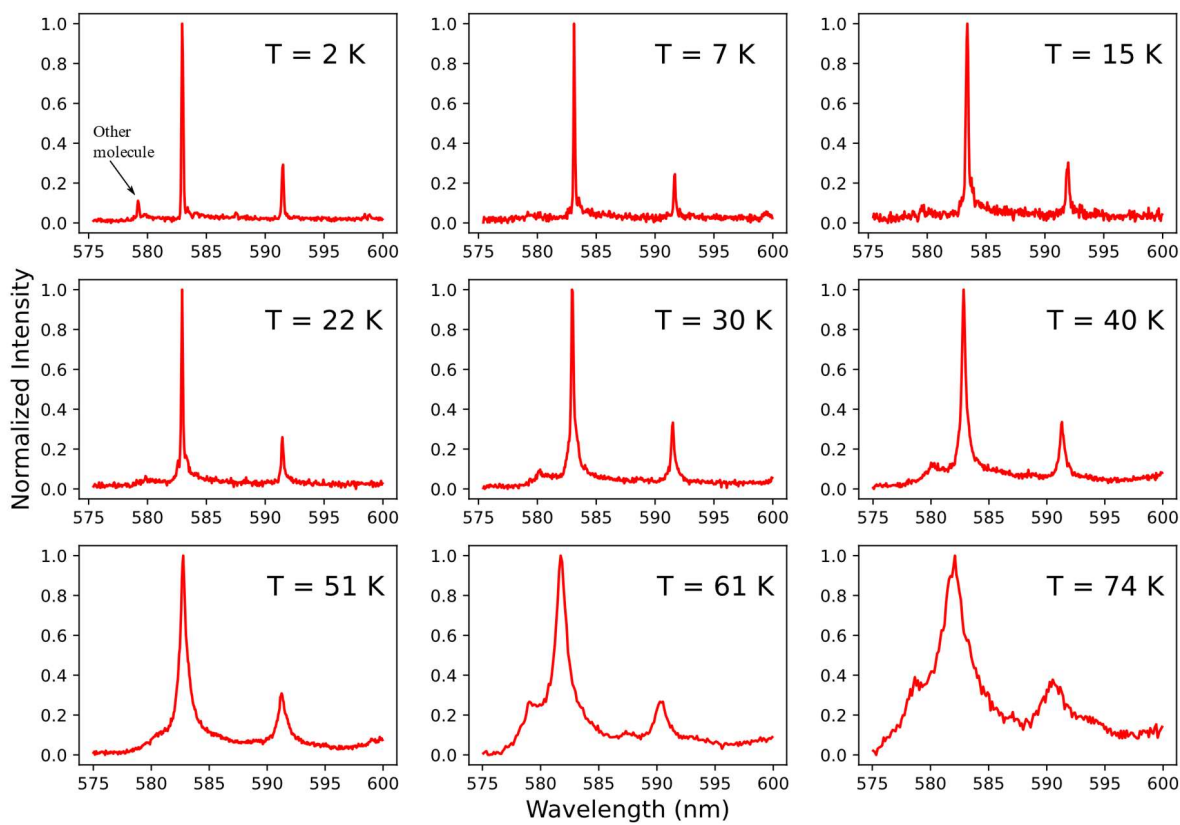

**Figure S4.1.** Nine emission spectra recorded at intermediate temperatures in a range from 2 K up to 74 K. From the main peak around 583 nm, we extract the linewidth from the full-width-at-half-maximum, in order to build up Figure 2b in the main text. A weak line of another emitter is present around 579 nm (indicated by arrow in top left figure). The first five spectra were integrated for 10 seconds, while the remaining spectra were integrated for 30 seconds. As the spectral lines broaden with temperature, the fluorescence intensity will be distributed over more pixels. Hence, with a longer integration we still obtained a relatively good signal per pixel.

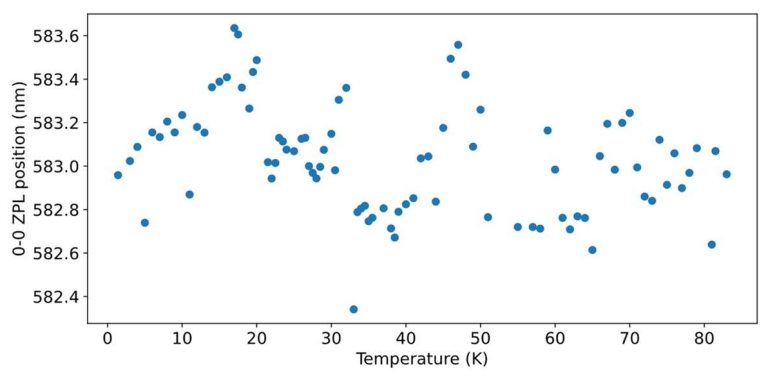

**Figure S4.2.** 0-0 ZPL position with temperature, determined from the 88 spectra that built up Figure 3b in the main text. The spectral position of the 0-0 ZPL may vary due to spectral jumps, which are indeed present as observed in Figure S3.3. Overall, the 0-0 ZPL position shows no temperature-induced shift or trend in the measured temperature range.

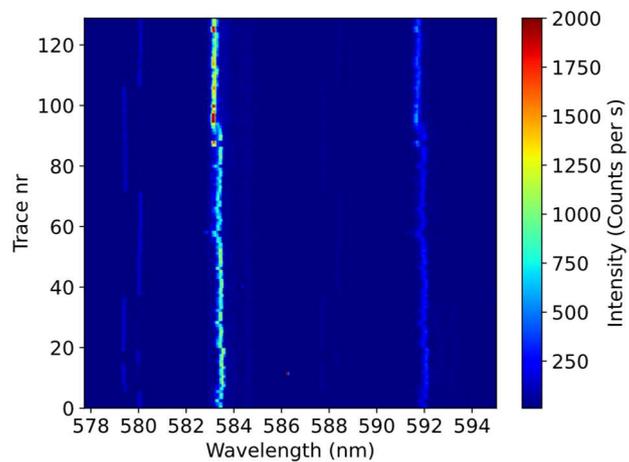

**Figure S4.3.** Time series of spectra of the molecule that we measured for the broadening of the 0-0 ZPL with temperature in Figure 2b in the main text. There are spectral jumps present that we attribute to the fluctuations in the main 0-0 ZPL position in Figure S4.2. Possibly, these fluctuations may add uncertainty to the points in Figure 2b in the main text, as the relatively long integration time of 10-30 s per spectrum may broaden the 0-0 ZPL by spectral jumps occurring within that time frame. The emitter appearing around 579-580 nm, as observed for the spectra in Figure S4.1, is also present in this series and shows clear coupling to a two-level system. T = 2 K.

## S5 Spectral wandering

Spectral instabilities of the single molecules are always present on our samples, in the form of clear jumps between two levels up to very complex jittering of the spectral position of the molecule. In this section we will show results of experiments we performed to characterize and find the source of the spectral diffusion.

*Photo-induced spectral jumps*

As we describe in the main text, the spectral diffusion is photo-induced at constant temperature. We confirmed this by running a series of spectra, while closing the shutter for two short periods (Figure S5.1). After the shutter was opened again, the molecules continued the trace at the position where they left.

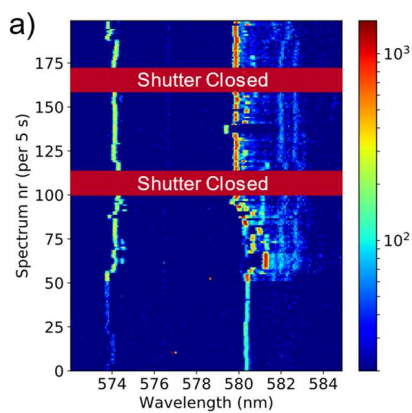

**Figure S5.1.** Spectral time trace of the 0-0 zero-phonon lines of a couple of molecules that are in the focal spot. For the first 50 traces the excitation intensity is 6 kW/cm$^2$ and the remainder is recorded with a 60 kW/cm$^2$ excitation intensity: the excitation intensity was increased to have a faster rate of spectral jumps and increase the number of visible molecules. The red regions indicate periods during which the shutter for the excitation path was closed. Once the shutter is re-opened, the molecules indeed continue from the same spectral position they had before the shutter was closed. T = 2 K.

*Spatial diffusion*

As hBN is known to be an excellent lubricant, we considered that the molecule might 'skate' over the surface at room temperature, or perhaps even at low temperature: spatial diffusion. Such a spatial diffusion might involve rotations of the molecule, which we can explicitly track by following the polarization of the molecule's fluorescence. The polarization of fluorescence is determined by the transition dipole moment vector, which for terrylene is oriented along the long axis of the molecule. To track changes in the polarization of the fluorescence, we installed a polarizing beam splitter in our setup, to separate the s-polarization and p-polarization components. We followed both polarization channels at the outputs of the beam splitter by recording the fluorescence signals with an APD. Simultaneously, we recorded spectra of the molecule, in order to detect spectral jumps. When a molecule rotates, the signals are expected to redistribute over the two detectors, i.e. show anticorrelated fluctuations. In Figure S5.2 we trace the fluorescence signal of the two polarization channels, with corresponding spectra of the molecule on the right side. For the second molecule in particular (c, d), there is a clear correlation between the two channels. Hence, the spectral jumps are not caused by rotation of the molecule.

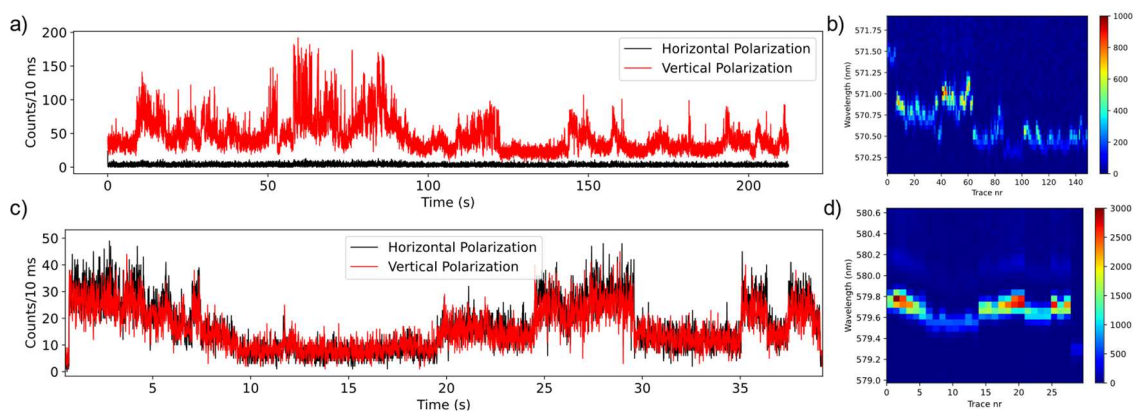

**Figure S5.2.** a,b) Fluorescence time trace and spectra of molecule A, which is nearly perfectly oriented with the vertical polarization component. Spectral jumps during the time trace have no influence on the horizontal polarization component. c,d) Shows the time trace and spectra of another molecule (molecule B) with comparable intensities in both polarization channels. Also here the spectral jumps have no influence on the ratio of the two polarization components. Around 28 s, molecule B jumps to a spectral position where the excitation was much less efficient, the signal dropped considerably, and therefore the data was cut here. The spectra in b) and d) were recorded with a 1 s integration time. The time scales of the fluorescence time trace and of the spectra are different, due to some delay between the recorded spectra. T = 2 K.

Even though there is no rotational diffusion, there is still a possibility of a translational motion. Below the diffraction-limit of our setup, translational motion could be detected using super-resolution imaging. As the molecule itself is a point source, compared to a diffraction-limited focal point, the position of the molecule can be deconvoluted by fitting the point-spread function (PSF) of the molecule by a 2D Gaussian of the form:

$$G(x,y) = B + A * e^{-((x-x_0)^2 + (y-y_0)^2)/\Sigma} \quad (1).$$

Here, $B$ is defined as the background, $A$ the amplitude of the PSF, $x_0$ and $y_0$ the coordinates of the molecule in the focal plane and $\Sigma$ a parameter characterising the spread of the PSF, assumed to be equal for the x and y direction. To measure the PSF of the molecule, we scanned the laser beam over the imaging plane. With more than two molecules present in the scan, we can trace the average distance between them by fitting both PSFs of the molecules to equation 1. An example of the experimental data and fit is shown in Figure S5.3. Both the integration time and the pixel width will influence the resolution of the molecule localization. To increase resolution, we programmed a 50 nm pixel size, with an integration time of 20 ms per pixel. Despite the high resolution of the scan and relatively high number of counts, the localization will be influenced by fluctuations of the fluorescence signal caused by spectral jumps.

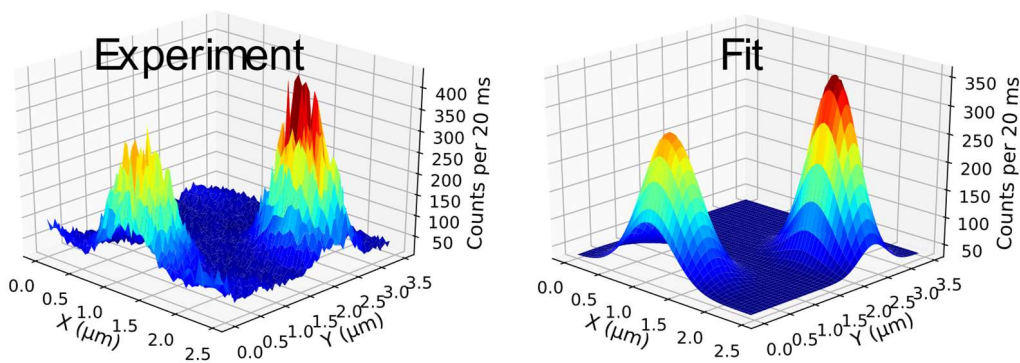

**Figure S5.3.** Point-spread functions of two molecules that are approximately 2 µm apart from each other and do not (or barely do) overlap with the PSFs from other molecules. The left image shows the experimental data and the right image is the fit to the experimental data. A single scan consisted of 100x100 pixels over a range of 5x5 µm² and took about 200 s to acquire with the scanning mirror. In total, 24 of these images were acquired over the course of 90 minutes with short pauses in between. T = 2 K.

The relative distance of the two molecules in the set of images taken over a period of 90 minutes, as deviations from the mean distance, are shown in Figure S5.4. Despite the relatively high number of counts, the resolution may be in particular limited due to spectral jumps. With a fitted FWHM (full width at half maximum) of approximately 700-800 nm (about 2.4 standard deviations of the Gaussian) and a total of 30-60 thousand counts per PSF, the expected resolution would be at best $\frac{700}{\sqrt{60000}} \approx 3\ nm$, which is overall the error obtained from the fit. Indeed, the difference between each measurement is much larger than 3 nm, but no obvious trend or drift in the position appears. The recorded distance between the molecules reverts around the mean, whereas a random walk would lead to an increase of the distance. Hence, if there is any spatial diffusion, it is likely very limited, despite the many spectral jumps that occurred over the long measurement time.

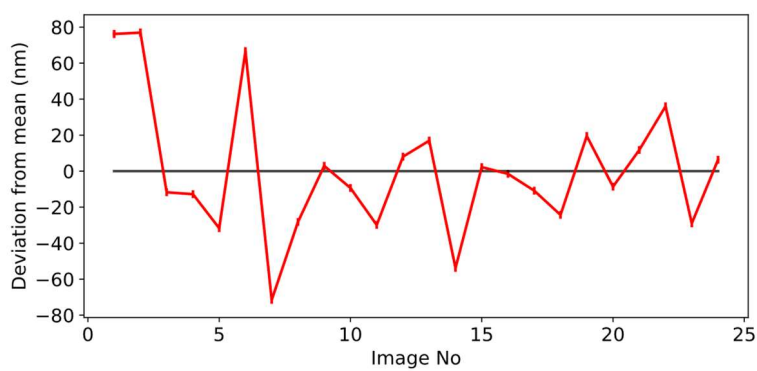

**Figure S5.4.** Variations of the distance between the two molecules of Fig. S5.3 extracted from 24 consecutive confocal images. The average error of the fit is just a few nm, but the measurements likely have a larger error due to spectral jumps occurring while recording the PSF. For all fitted data, the distance appears to revert around a mean value of 2.18 µm, and shows no clear trend. Hence, if there is any spatial diffusion, its extent is very limited. T = 2 K.

*The substrate*

Although our hBN flakes are relatively thick, as determined by AFM in section 1, we considered that the substrate could play a role for the spectral jumps. For some defects inside hBN, albeit for relatively thin flakes of 13 nm, the spectral stability has been reported to improve by coating the substrate with an alumina layer, to passivate the silica[14]. On one of our samples we found a flake that did not completely attach to the substrate and a part of it was free-standing at a large angle (Figure S5.6). Terrylene molecules had still condensed on the surface by vacuum sublimation, as can be observed on the part that is in focus in Figure S5.5a. The spectral time traces we recorded of two bright spots show that there are still spectral jumps present.

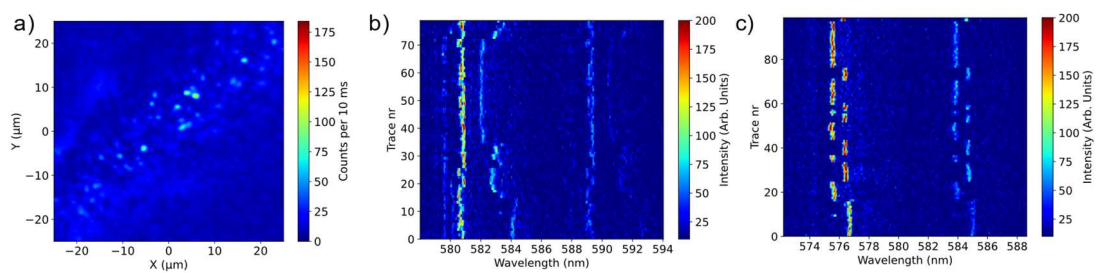

**Figure S5.5**. a) Shows a fluorescence image from the part of an hBN flake that is not completely attached to the silicon/silica substrate, but was standing free at an angle with the substrate. We deposited terrylene molecules on this flake by vacuum sublimation. Only a part of the flake is in focus with the laser beam, along the diagonal of the figure. b) and c) show spectral traces of some terrylene molecules that are in focus, with their 0-0 ZPL and their strongest vibrational line around 248 cm$^{-1}$. Although there is no substrate in the vicinity of these molecules, large spectral jumps are still observed. The spectrum in b) is measured at the two dots in a) at coordinates (5.4, 8.0) and the spectrum in c) is measured at the single dot at coordinates (-6,-4) in a). T = 2 K.

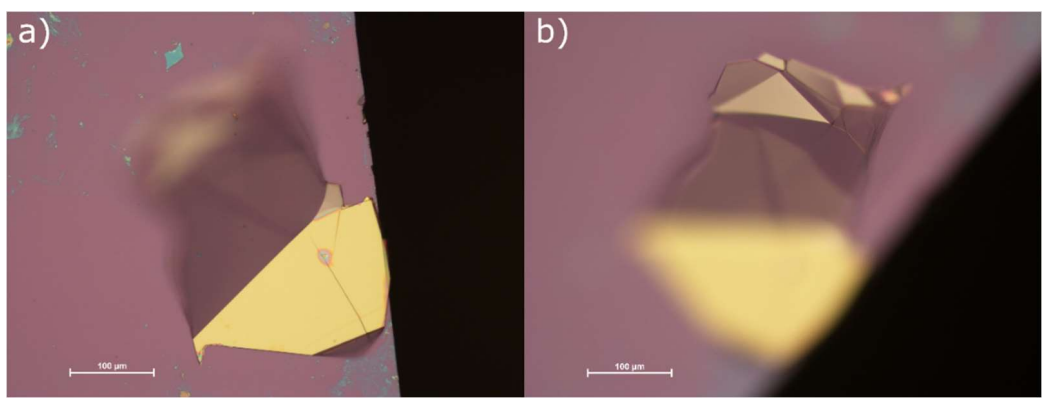

**Figure S5.6.** a) Microscope image of the flake that is measured for Figure S5.5. The part in focus is in contact with the substrate, while in b) the non-attached part is in focus, above the surface of the substrate

*Annealing*

For single defects in CVD grown hBN it is reported that the width of the ZPL at room temperature was significantly improved by thermally annealing the hBN, with a clear improvement between an annealing temperature of 550 ˚C and 750 ˚C[15]. Similarly, we find that the spectral stability of the molecules improve significantly by annealing the flakes prior to deposition of the molecules. At 750 ˚C, maintained overnight for 12 hours, we obtain the best results. A subset of series of spectra are shown in Figure S5.7, which are complementary to the spectra shown in Figure 4 in the main text.

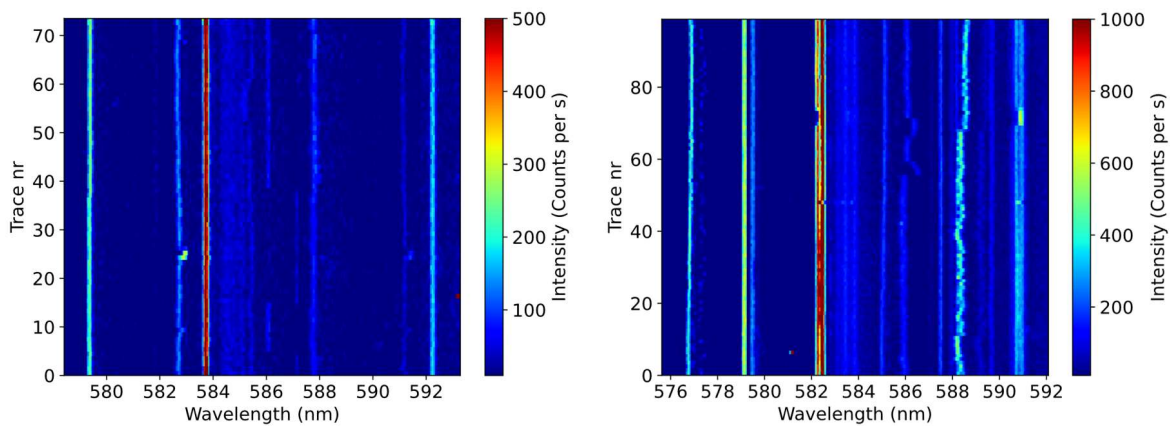

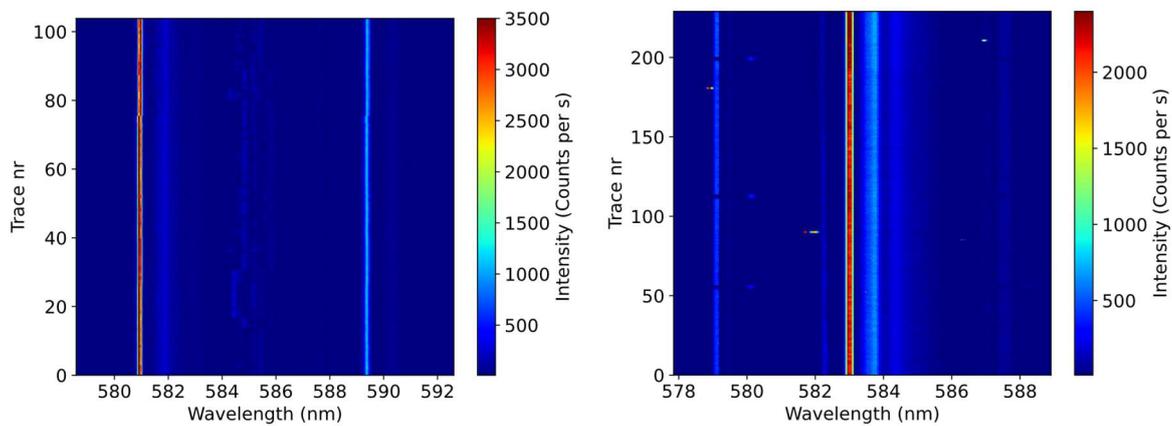

**Figure S5.7**. Additional spectral series of annealed samples (750 °C for 12 hours). Some molecules are more isolated than others, but even in a small ensemble, as in the top right figure, most emitters do not move much. T = 2 K.

## S6 Photon antibunching

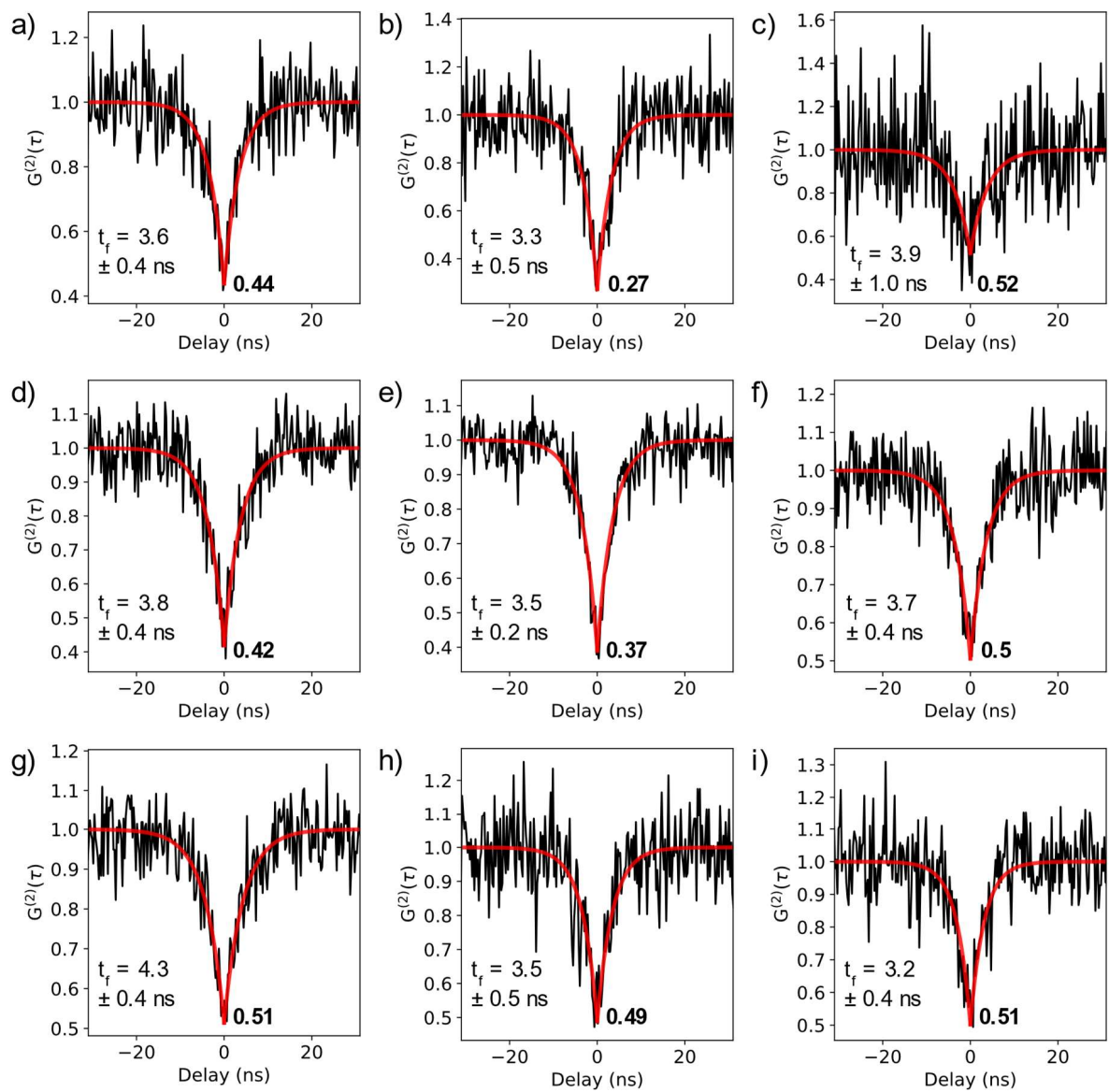

**Figure S6.1.** Collection of antibunching measurements on the fluorescence signal of terrylene. Four out of nine antibunching curves display a contrast of less than 50 %, which in all cases is consistent with the measured background signal. The lifetimes obtained from the fit are displayed on the bottom left and give an average of 3.6 ± 0.2 ns. The antibunching curve displayed in e) is also shown in the main text. The spectra of these nine molecules were measured as well, to confirm that the spectral signature corresponded to terrylene. T = 2 K.

## S7 Low-temperature excitation spectra of the 0-0 ZPL

We use a tunable single-frequency laser, with a linewidth of about 1 MHz, to record excitation spectra of the 0-0 ZPL of single molecules. To find molecules, we position our laser's frequency around the peak of the inhomogeneous broadening, somewhere in between 581 and 583 nm. Furthermore, we set the laser intensity to be high enough to power-broaden 0-0 ZPLs in order to increase the number of molecules that emit fluorescence in a spectral range around the laser's frequency. Otherwise, the number of responsive molecules would be very low, due to the low concentration of molecules and the relatively large inhomogeneous broadening. To observe the responsive molecules, we record a fluorescence map of the flakes. An example is shown in Figure S7.1a. The bright pixels around (20, -25) belong to the molecule in Figure 5 in the main text.

Without positioning our laser on the molecule directly, we reduce the laser intensity to levels of a few W/cm$^2$ and start a scan of the laser. We do this to prevent the molecule from jumping away before we even start scanning the laser. After the first scan is completed, we move the laser spot to the position of the molecule. About half the time, we observe molecules with similar behaviour as shown in Figure S7.1b. A strong spectral diffusion, possibly in combination with spectral jumps, makes the molecule appear intermittently over the full scan range. From these type of molecules it is impossible to deduce a linewidth.

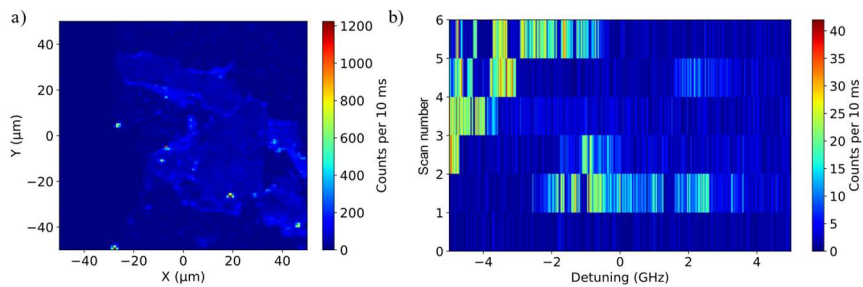

**Figure S7.1**. a) Fluorescence map of hBN crystals at an excitation wavelength of 582.38 nm and a laser intensity of 350 W/cm$^2$. The bright spot around (20, -25) belongs to the molecule of Figure 5 in the main text. b) shows excitation spectra of a single molecule with very common behaviour for the 0-0 ZPL. The strong spectral diffusion and blinking makes it difficult to extract a linewidth. The sample was annealed at 750 ˚C for 12 hours and measurement temperature was 2 K.

For the remainder of the molecules it was possible to observe the 0-0 ZPL and fit a Lorentzian distribution to the fluorescence profile, as was the case for Figure S7.2d and Figure S7.3b. In some cases, a Gaussian distribution fits better, such as Figure S7.4b. In general, molecules can be followed for a limited time period, after which they jump out of the scan range. This occurs for the molecules in Figure S7.3a and S7.4a. Discrete jumps can also be observed within the scan range, such as in Figure S7.3a. This molecule has two Lorentzian distributions, spaced by a gap of 1.1 GHz. The two distributions have the same linewidth and the spacing remains intact after a spectral jump. Hence, the two lines are likely from the same molecule, which may be coupling to a fast-switching two-level system responsible for the line splitting.

In addition to discrete spectral jumps, molecules also show jittering spectral diffusion, which becomes stronger as the laser intensity is increased. Figure S7.2a,b and c show examples of such photo-induced spectral diffusion. Apart from expected power broadening of the linewidth, the amplitude of the spectral diffusion clearly increases as the laser intensity is raised by a factor 10 between the time series from a to c. Despite the spectral diffusion, the molecule was never lost in our limited scanning range of 10-20 GHz. Therefore we can do time-correlated single-photon counting and observe antibunching in the fluorescence emission (Figure S7.2e). As the molecule is driven resonantly by the laser, Rabi oscillations are expected to show up at these high excitation intensities. However, they are not clearly present, which means there is probably significant dephasing present or they are blurred by spectral diffusion. Despite absence of spectral diffusion for the molecule in Figure 5 in the main text, no Rabi oscillations appear in the antibunching histogram as well (Figure S7.5).

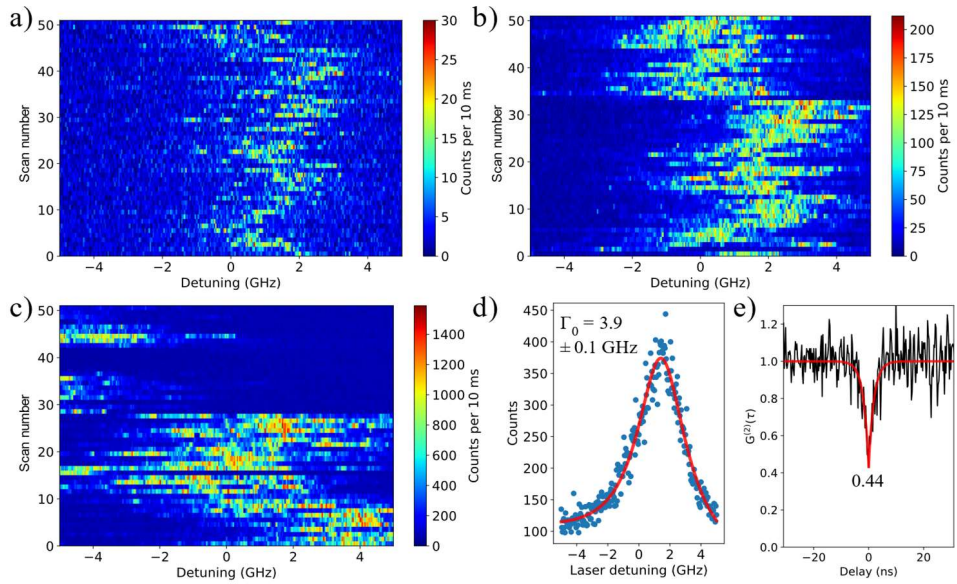

**Figure S7.2.** a) Excitation spectra of the 0-0 ZPL of a single terrylene molecule measured at three different excitation intensities. The excitation intensity was 1 W/cm$^2$ in a), 14 W/cm$^2$ in b) and 250 W/cm$^2$ in c). d) We extracted the linewidth of the emitter by summing all excitation spectra in a), fitted to a Lorentzian distribution. In e) we recorded the correlation function of the resonance fluorescence, showing a dip at zero delay. The antibunching histogram was recorded with the high excitation power in c), which led to long blinking of the fluorescence signal due to spectral diffusion. The blinking, residual background and possibly limits of the instrument as a result of the short decay time, attributed to the relatively high dip at 0.44. We observe no Rabi oscillations, which is likely due to significant dephasing or spectral diffusion. The short decay time is likely caused by stimulated emission. The sample was annealed at 750 °C for 12 hours and measurement temperature was 2 K.

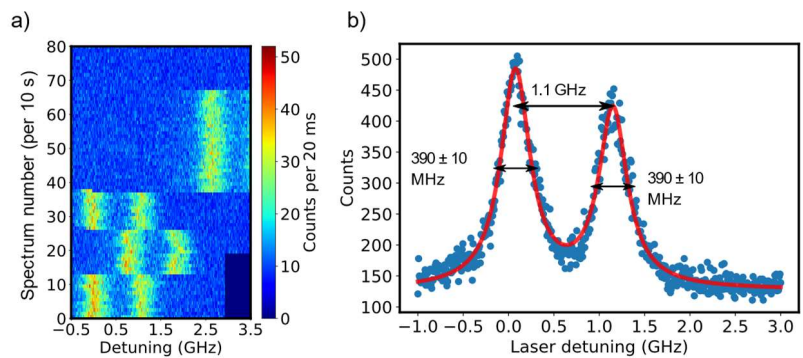

**Figure S7.3.** a) Series of emission spectra of a molecule with a relatively narrow linewidth, found on a non-annealed flake. The two lines are likely related to coupling to a fast-switching two-level system, while slower jumps are related to slow-switching two-level systems. After about 10 minutes, the molecule jumped out of the scanned range. The complete series is composed of two individual scans that were shifted with respect to each other. The gap in the scan on the bottom right is the result of the shifted position of the two individual scans. b) shows the sum of the first ten excitation spectra, which is fit to a sum of two Lorentzian distributions. The linewidths of the two fits give the same number of 390 ± 10 MHz, about 8-9 times larger than the lifetime-limited linewidth.

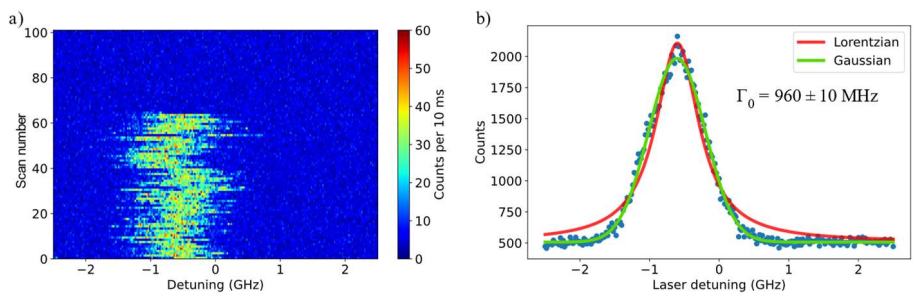

**Figure S7.4.** a) Series of excitation spectra of a molecule at 581.96 nm, excited with a laser intensity of 11 W/cm$^2$. The molecule shows weak spectral diffusion, but jumps out of the scanned range at scan 65. b) Excitation spectrum of the 0-0 ZPL, which is derived from the sum of all excitation spectra in a). Due to the short tails, the spectral profile fits best to a Gaussian distribution (green line) with a linewidth of 960 ± 10 MHz. The red line shows the corresponding fit of a Lorentzian distribution. The sample was annealed at 750 °C for 12 hours and measurement temperature was 2 K.

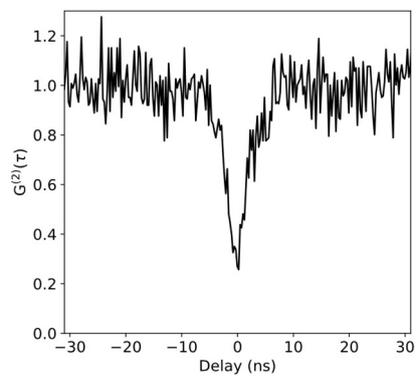

**Figure S7.5.** Antibunching histogram for the resonantly excited molecule of Figure 5 in the main text, measured at 2 K. There is no clear presence of Rabi oscillations, while the laser intensity of 285 W/cm$^2$ was considerably higher than the saturation intensity of 105 ± 22 W/cm$^2$. As the linewidth is about 4.7 ± 0.3 GHz, the coherence time would be a minimum of 68 ± 4 ps, which is too short to be detected with our setup. The characteristic time of the exponential increase, starting from zero delay, of about 2.5 ns is significantly shorter than found for non-resonant excitation. This is likely the result of stimulated emission due to the relatively high intensity of the laser.

## S8 Impurity emitters on hBN

We set up a control experiment with hBN flakes that were spin-coated with toluene with no dye dissolved or flakes which have undergone no treatment at all. On the untreated hBN flakes no narrow emitters were found. However, on the flakes that were spin-coated with toluene several narrow emitters could be found. Frequently occurring emitters clearly have the vibrational fingerprint (Figure S8.1) of a well-known impurity that is typically found in organic and polymer layers[16,17], sometimes denoted as molecule X. In PMMA matrices these emitters were found over a broad range of 1.9 eV (650 nm) up to 2.2 eV (570 nm). On hBN we find spectra characteristic of molecule X or compounds of its family over a wavelength range from 618 nm up to 640 nm (Figure S8.2), which is similar as the distribution found in polymers[17].

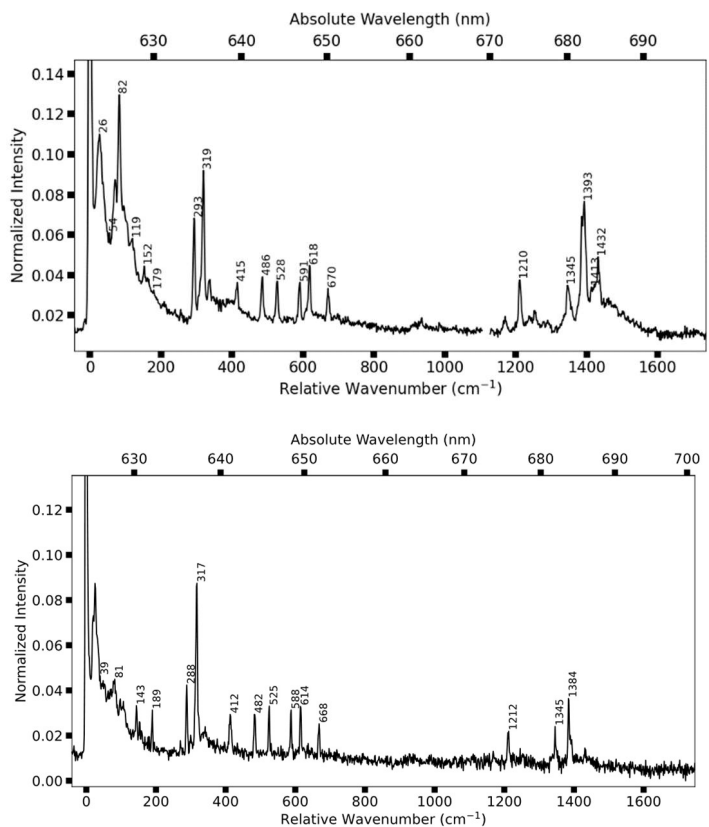

**Figure S8.1.** Two spectra of the impurity molecules that we find on samples prepared by spin-coating with a toluene solution. Some of the peak's intensities vary significantly from emitter to emitter.

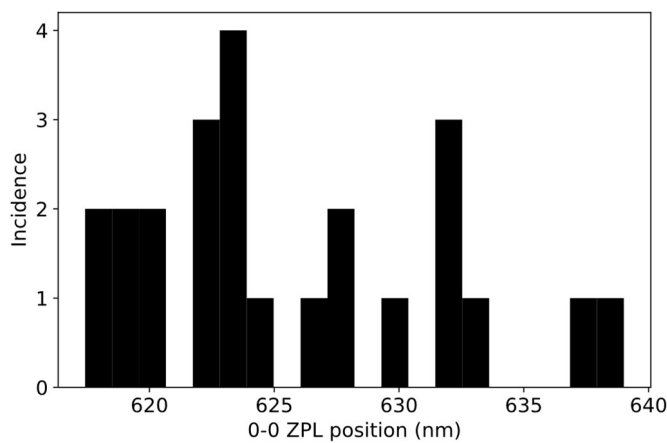

**Figure S8.2**. Histogram of the 0-0 ZPL positions of the same impurity emitters as in Figure S6.1. A total of 24 of these emitters were found in our experiments.

**References**

1. Zhang, Y. *et al.* Probing Carrier Transport and Structure-Property Relationship of Highly Ordered Organic Semiconductors at the Two-Dimensional Limit. *Phys. Rev. Lett.* **116**, 016602 (2016).
2. Park, B. *et al.* Anomalous Ambipolar Transport of Organic Semiconducting Crystals via Control of Molecular Packing Structures. *ACS Appl. Mater. Interfaces* **9**, 27839–27846 (2017).
3. Palewska, K. *et al.* Total Luminescence Spectroscopy of Terrylene in Low-Temperature Shpol'skii Matrixes. *J. Phys. Chem.* **99**, 16835–16841 (1995).
4. Deperasińska, I. & Kozankiewicz, B. Non-planar distortion of terrylene molecules in a naphthalene crystal. *Chem. Phys. Lett.* **684**, 208–211 (2017).
5. Kurdyumov, A. V., Solozhenko, V. L. & Zelyavski, W. B. Lattice Parameters of Boron Nitride Polymorphous Modifications as a Function of Their Crystal-Structure Perfection. *J. Appl. Crystallogr.* **28**, 540–545 (1995).
6. Deperasińska, I., Kozankiewicz, B., Biktchantaev, I. & Sepioł, J. Anomalous Fluorescence of Terrylene in Neon Matrix. *J. Phys. Chem. A* **105**, 810–814 (2001).
7. Hod, O. Graphite and Hexagonal Boron-Nitride have the Same Interlayer Distance. Why? *J. Chem. Theory Comput.* **8**, 1360–1369 (2012).
8. Kim, S. M. *et al.* Synthesis of Patched or Stacked Graphene and hBN Flakes: A Route to Hybrid Structure Discovery. *Nano Lett.* **13**, 933–941 (2013).
9. Lee, W. H. *et al.* Surface-Directed Molecular Assembly of Pentacene on Monolayer Graphene for High-Performance Organic Transistors. *J. Am. Chem. Soc.* **133**, 4447–4454 (2011).
10. Cassabois, G., Valvin, P. & Gil, B. Hexagonal boron nitride is an indirect bandgap semiconductor. *Nat. Photonics* **10**, 262–266 (2016).
11. Navarro, P. *et al.* Electron Energy Loss of Terrylene Deposited on Au(111): Vibrational and Electronic Spectroscopy. *J. Phys. Chem. C* **119**, 277–283 (2015).
12. Tchénio, P., Myers, A. B. & Moerner, W. E. Optical studies of single terrylene molecules in polyethylene. *J. Lumin.* **56**, 1–14 (1993).
13. Nicolet, A., Kol'chenko, M. A., Kozankiewicz, B. & Orrit, M. Intermolecular intersystem crossing in single-molecule spectroscopy: Terrylene in anthracene crystal. *J. Chem. Phys.* **124**, 164711 (2006).
14. Li, X. *et al.* Nonmagnetic Quantum Emitters in Boron Nitride with Ultranarrow and Sideband-Free Emission Spectra. *ACS Nano* **11**, 6652–6660 (2017).
15. Li, C. *et al.* Purification of single-photon emission from hBN using post-processing treatments. *Nanophotonics* **8**, 2049–2055 (2019).
16. Fleury, L. *et al.* Single Molecule Spectra of an Impurity Found in N-Hexadecane and Polyethylene. *Mol. Cryst. Liq. Cryst. Sci. Technol. Sect. Mol. Cryst. Liq. Cryst.* **283**, 81–87 (1996).
17. Neumann, A., Lindlau, J., Thoms, S., Basché, T. & Högele, A. Accidental Contamination of Substrates and Polymer Films by Organic Quantum Emitters. *Nano Lett.* **19**, 3207–3213 (2019).